\documentclass[nofootinbib,preprintnumbers,prd,superscriptaddress,showpacs,twocolumn]{revtex4}

\usepackage[titletoc]{appendix}
\usepackage{color}
\usepackage{hyperref}

\usepackage[rightcaption]{sidecap}
\usepackage{subfigure}
\usepackage{comment}

\usepackage{graphicx}
\usepackage{amsmath}
\usepackage{amsfonts}
\usepackage{amsthm}
\usepackage{mathrsfs}
\usepackage{amssymb}
\usepackage{epsfig}
\usepackage{amscd}
\usepackage{dcolumn}
\usepackage{bm}
\usepackage{natbib}
\usepackage{url}
\usepackage{xspace}

\usepackage{dcolumn}

\usepackage{array}
\usepackage{ctable}
\usepackage{multirow}
\usepackage{siunitx}
\usepackage{longtable}
\usepackage{tabularx}
\usepackage{booktabs}

\usepackage{amsthm}
\usepackage{mathrsfs}

\usepackage{epsfig}
\usepackage{amscd}

\usepackage{bm}
\usepackage{natbib}
\usepackage{url}
\usepackage{xspace}
\usepackage{kantlipsum}

\graphicspath{{Graphics/}}

\def\be{\begin{equation}}
\def\ee{\end{equation}}
\def\bea{\begin{eqnarray}}
\def\eea{\end{eqnarray}}

\def\nat{Nature}

\def\prd{Phys. Rev. D}

\def\mnras{MNRAS}
\def\aj{AJ}
\def\apj{ApJ}
\def\apjl{ApJ Lett.}
\def\apjs{ApJ Suppl. Ser.}

\def\aap{A\&A}

\def\pasj{Publ. Astr. Soc. Japan }

\def\physrep{Phys. Rep.}
\def\jcap{JCAP}

\definecolor{vividviolet}{rgb}{0.62, 0.0, 1.0}
\definecolor{amaranth}{rgb}{0.9, 0.17, 0.31}
\definecolor{palatinateblue}{rgb}{0.15, 0.23, 0.89}
\definecolor{brightpink}{rgb}{1.0, 0.0, 0.5}
\definecolor{cornflowerblue}{rgb}{0.39, 0.58, 0.93}
\definecolor{deepcarminepink}{rgb}{0.94, 0.19, 0.22}
\definecolor{radicalred}{rgb}{1.0, 0.21, 0.37}

\hypersetup{ linktoc=all,
    colorlinks, linkcolor={palatinateblue},
    citecolor={brightpink}, urlcolor={amaranth}
}

\begin{document}

\title{Characterizing the equivalence between dark energy and radiation using gamma-ray bursts}

\author{Orlando Luongo}
\email{orlando.luongo@unicam.it}
\affiliation{University of Camerino, Via Madonna delle Carceri, Camerino, 62032, Italy.}
\affiliation{SUNY Polytechnic Institute, 13502 Utica, New York, USA.}
\affiliation{INAF - Osservatorio Astronomico di Brera, Milano, Italy.}
\affiliation{Istituto Nazionale di Fisica Nucleare (INFN), Sezione di Perugia, Perugia, 06123, Italy.}
\affiliation{Al-Farabi Kazakh National University, Al-Farabi av. 71, 050040 Almaty, Kazakhstan.}

\author{Marco Muccino}
\email{marco.muccino@lnf.infn.it}
\affiliation{University of Camerino, Via Madonna delle Carceri, Camerino, 62032, Italy.}
\affiliation{Al-Farabi Kazakh National University, Al-Farabi av. 71, 050040 Almaty, Kazakhstan.}
\affiliation{ICRANet, Piazza della Repubblica 10, 65122 Pescara, Italy.}

\begin{abstract}
Differently from the equivalence time between either matter and radiation or dark energy and matter, the equivalence between dark energy and radiation occurs between two subdominant fluids, since it takes place in the matter dominated epoch. However, dark energy--radiation equivalence may correspond to a \emph{cosmographic bound} since it  strongly depends on how dark energy evolves. Accordingly, a possible model-independent bound on this time would give hints on how dark energy evolves in time. In this respect, gamma-ray bursts (GRBs) may be used, in fact, as tracers to obtain cosmic constraints on this equivalence. Consequently, based on observed GR data from the $E_{\rm p}$--$E_{\rm iso}$ correlation, we here go beyond by simulating additional GRB data points and investigating two distinct equivalence epochs: 1)  dark energy--radiation, and 2) dark energy--radiation with matter. We thus extract constraints on the corresponding two redshifts adopting Monte Carlo Markov chain simulations by means of two methods: the first performing the GRB calibration and the cosmological fit steps independently, and the second performing these steps simultaneously by resorting a hierarchical Bayesian regression. To keep the analysis model-independent, we consider a generic dark energy model, with the unique constraint to  reduce to the $\Lambda$CDM at $z=0$. Our
findings are thus compared to theoretical predictions, indicating that the $\Lambda$CDM model is statistically favored to predict such an equivalence time, though a slow evolution with time cannot be fully excluded. Finally, we critically re-examine the Hubble constant tension in view of our outcomes.
\end{abstract}


\maketitle
\tableofcontents

\section{Introduction}
\label{intro}

Dark energy and dark matter constitute the majority of the energy budget of our Universe \cite{Planck2018} and their fundamental nature is still object of heated  debate\footnote{More precisely, dark energy is responsible for present-time cosmic acceleration \cite{1998Natur.391...51P,1998AJ....116.1009R,1999ApJ...517..565P,2003ApJ...594....1T,2003Sci...299.1532B,2003ApJS..148....1B,2003ApJS..148..135H,2003ApJS..148..161K,2003ApJS..148..175S,2005ApJ...633..560E}, exhibiting repulsive effects \cite{Luongo:2014qoa}, while dark matter plays a crucial role in cosmic structure clustering \cite{Bergstrom:2009ib,Profumo:2019ujg}.}.

In the quest to understand dark energy, various hypotheses have been put forth, all seeking to determine whether its equation of state changes with the time  \cite{Chevallier2001,Linder2003,2003RvMP...75..559P,King2014}, or if it remains a pure cosmological constant $\Lambda$ \cite{2003PhR...380..235P}.
The latter is generally associated with the effects of primordial quantum fluctuations and represents the key ingredient of the \emph{standard background scenario}, namely the $\Lambda$CDM model \cite{2000IJMPD...9..373S,2006IJMPD..15.1753C,tsujikawa2011dark}, essentially based on six free parameters \citep{Planck2018,2021arXiv210505208P}, with the further assumption of flat topology, as supported by various observations\footnote{The precise determination of spatial curvature $\Omega_k$ still remains an open challenge in cosmology \cite{2018ApJ...864...80O}.} \citep{2020MNRAS.496L..91E}.

For its theoretical structure, the $\Lambda$CDM model seems to be statistically favored to frame the dark energy dynamics, especially at late and early times. However, cosmological tensions and conceptual issues have been recently raised\footnote{For different perspectives about extensions of the standard model, see e.g. Refs.~\citep{nostro,2022CQGra..39s5014D,mio2022}.} \cite{Hu:2023jqc}, as well as evidence in favor of evolving dark energy even at late times \cite{2024arXiv240403002D}.

Consequently, the use of additional standard candles, such as type Ia supernovae (SNe Ia) \cite{2018ApJ...859..101S}, and/or standard rulers, such as baryonic acoustic oscillations (BAOs) \cite{2019JCAP...10..044C}, may be not enough to state whether dark energy is dynamical or not. Hence, intermediate- and high-redshift catalogs beyond SN Ia detectability \citep{Rodney2015} appear crucial to really understand whether the $\Lambda$CDM model may be seen as a limiting case of a more general paradigm \citep{2019IJMPD..2830016C,2020arXiv200309341C}.

Accordingly, GRBs are currently inspected as potential distance indicators \cite{2021ApJ...908..181M,2022MNRAS.512..439C,2021Galax...9...77L,2022PASJ..tmp...83D,2022arXiv220809272J}. Particularly, GRBs may crucially detect deviations from the well-established $\Lambda$CDM predictions, clarifying the nature of the aforementioned cosmological tensions.

In this work, we explore the potential use of GRBs to detect limits on the equivalence between dark energy and additional cosmic fluids. Precisely, while the equivalence between matter and radiation applies to very high redshifts, culminating into the \emph{equivalence epoch}, it is challenging to directly constrain either the equivalence between dark energy and radiation or dark energy and matter with radiation\footnote{It is worth to stress that matter--radiation equivalence represents a  \emph{cosmological epoch}, i.e., a time before which radiation dominates over other fluids, while equating dark energy with radiation or with radiation plus matter provides \emph{cosmographic redshifts}, without implying a time of domination of one of the two species.}. With existing data, these two equivalence times and their corresponding formal periods can be tentatively constrained and, since they are influenced by the dark energy form and free parameters, they can be used to \emph{directly} discriminate among dark energy models and investigate \emph{in a model-independent way} the possible dark energy evolution, representing \emph{de facto} a relatively unexplored subject in the literature. To pursue this goal, we consider a generic dark energy model that reduces to the $\Lambda$CDM paradigm at redshift $z=0$.
Afterwards, we fit our generic model and constrain the equivalence epochs against observed and simulated GRB data points up to very high redshifts, i.e., $z\simeq12$. To do so, we develop two \emph{model independent} methods: the first performs GRB  calibration and cosmological fits \emph{independently}, whereas the second involves a hierarchical Bayesian regression, i.e., performing GRB calibration and cosmological fits \emph{simultaneously}, without passing through the calibration procedure. To do so, we employ Monte Carlo Markov chain (MCMC) fits, utilizing the Metropolis–Hastings algorithm for parameter estimation. The underlying likelihood functions are accordingly maximized under the assumption of Gaussian-distributed errors, with a modified version of the \texttt{Wolfram Mathematica} code, originally presented in Ref.~\citep{2019PhRvD..99d3516A}, whose cosmological applications with GRBs have been extensively tested, see e.g. Refs.~\cite{2019MNRAS.486L..46A,LM2020,2023MNRAS.518.2247L,2023MNRAS.523.4938M}. Our findings on the dark energy--radiation equivalence time have been computed working out the well-consolidate $E_{\rm p}$--$E_{\rm iso}$ correlation of GRBs and indicate that, based on current data, we cannot definitively exclude that dark energy remains constant at very high redshift even at 1--$\sigma$ confidence level. However, the predictions of the standard cosmological model consistently fall within the 2--$\sigma$ confidence levels, thus certifying that the $\Lambda$CDM model may appear statistically favored even at $z\gtrsim1$. Thus, as a matter of comparison, the results on the equivalence time between dark energy--radiation plus matter, occurring at redshifts extremely close to our time, show the statistical significance of the standard $\Lambda$CDM background over generic dark energy model. To confirm this, we also test a $w$CDM model  in both the aforementioned equivalence time domains, thus finding no evident need of having an evolving  equation of state, $w\neq-1$. Concluding, our approach is novel since it  offers a  discerning method towards plausible constraints obtained from the cosmographic equivalence times at different redshift domains. Our findings seem to be in favor of the standard paradigm, matching the criticisms \cite{2024arXiv241104878A, 2024arXiv240802536A, 2024arXiv240412068C, 2024A&A...690A..40L} against the last developments in favor of a slightly evolving dark energy contribution shown by the DESI collaboration \cite{2024arXiv240403002D}. Finally, we critically re-examine the Hubble
constant tension in view of our outcomes, wondering whether our method can be used to acquire more information toward the existence of the $H_0$ tension itself.

The paper is structured as follows. In Sec.~\ref{sec:2}, we introduce our generic dark energy model, the main features of both model-independent techniques of reconstruction, and how to build up equivalence-time discriminators. Predictions on our priors for different models are thus reported. In Sec.~\ref{sec:3}, the cosmographic reconstructions of our equivalence redshifts have been reported. GRB data from the $E_{\rm p}$--$E_{\rm iso}$ correlation are reproduced in Sec.~\ref{sec:4}. In Sec.~\ref{sec:5}, our numerical results are inferred and critically reinterpreted, and their theoretical interpretation is therefore summarized in Sec.~\ref{sec:6}, whereas in Sec.~\ref{sec:7} we develop conclusions and perspectives of our work.

\section{Equivalence redshifts between dark energy and cosmic fluids}
\label{sec:2}

In view of the large-scale homogeneity and isotropy of the Universe, the cosmological principle can be reformulated invoking a maximally-symmetric spacetime, represented by the Friedmann-Robertson-Walker (FRW) line element that, in a spatially-flat configuration, reads  $\displaystyle{ds^2=dt^2-a(t)^2\left[dr^2+r^2 \left(d\theta^2+\sin^2{\theta}d\phi^2\right)\right]}$. Plugging the metric in Einstein's field equations yields the Friedmann equations,
\begin{subequations}
\begin{align}
\label{Fried1}
H^2 &= \frac{8\pi G}{3}\rho\,,\\
\dot{H}+H^2  &= -\frac{4\pi
G}{3}\left(\rho+3P\right)\,,\label{Fried2}
\end{align}
\end{subequations}
describing the cosmological dynamics, incorporated in the Hubble parameter $H\equiv\dot a/a$, through the energy-momentum components, made up by the total pressure $P$ and density $\rho$ of \emph{each single barotropic fluid}.

Thus, since the total density and pressure appear in the Friedmann equations, the individual subcomponents, constituting $\rho$ and $P$, cannot be easily \emph{disentangled}, leading \emph{de facto} to a severe \emph{degeneracy problem} \cite{Aviles:2011ak,vonMarttens:2019ixw}. This limitation hampers our  understanding of the Universe's dynamics since the crux of the problem lies in the degeneracy between mass and dark energy within the total energy density of the Universe.

To mitigate this issue, we can focus on late and intermediate cosmic evolution, where the total equation of state may be mainly computed through  matter, radiation and dark energy contributions.

Consequently, the need of model-independent techniques for cosmological reconstructions, able to disclose information on the Hubble rate derivatives, appear essential  \cite{Luongo:2022bju,Aviles:2016wel,Izzo:2010ix,Aviles:2012ay,Dunsby:2015ers}.
Phrasing it differently, cosmological treatments that do not fix the Hubble rate \emph{a priori}, formulating a direct model-independent dark energy reconstruction, turn out to be quite important to characterize dark energy at intermediate redshifts \cite{2020JCAP...04..043L,2005GReGr..37.1555T}.

\subsection{Model-independent dark energy reconstruction}

In view of the considerations made above, limiting to matter, radiation and dark energy, the first Friedmann equation acquires the form \cite{Luongo:2015zgq,Capozziello:2022jbw}
\begin{equation}
\label{hz}
H(z)=H_0\sqrt{\Omega_{m}(1+z)^{3}+\Omega_r(1+z)^4+\Omega_{DE}G(z)},
\end{equation}
where $\Omega_{m}\equiv\rho_m/\rho_c$,  $\Omega_{r}\equiv\rho_r/\rho_c$ and $\Omega_{DE}\equiv\rho_{DE}/\rho_c$, in which $\rho_c\equiv 3H_0^2/(8\pi G)$ is  the current value of the critical density and $H_0=H(z=0)$ is the Hubble constant.

More precisely, in Eq.~\eqref{hz}, $G(z)$ is a generic dark energy density, of which we do not explicitly specify the form, and we treat matter and radiation as barotropic fluids, i.e., starting from their equations of state, $\omega_m=0$ and $\omega_r=1/3$, respectively\footnote{For a different perspective, involving an effective matter fluid with non-zero pressure, consider Ref.~\cite{2018PhRvD..98j3520L,2024PDU....4401458B,2024arXiv241111130B}.}.

To fulfill cosmological constraints, we notably opt for the simplest requirements \cite{Luongo:2015zgq}
\begin{equation}\label{condizioni}
\left\{
\begin{array}{llll}
G(z)= 1 &, & \hbox{$z=0$} &, \\
\,\\
\Omega_{DE}\equiv1-\Omega_{m}-\Omega_r &, & \hbox{$\forall z$} &,\\
\,\\
\Omega_{DE} G(z) \gtrsim \Omega_{m} (1+z)^3 &, & \hbox{$z\rightarrow 0$} &,\\
\,\\
\Omega_{DE}G(z) \gtrsim \Omega_{r} (1+z)^4 &, & \hbox{$z\rightarrow 0$} &,
\end{array}
\right.
\end{equation}
where the second one may directly follow from the first one, whereas the last two properties indicate that at late times dark energy dominates over matter and radiation.

In particular, the latter suggests that, if a cosmological equivalence between dark energy and radiation/matter magnitudes exists, then \emph{it might be present at intermediate epochs}, rather than at current time\footnote{Around current times, instead, a matter--dark energy equivalence may occur, see Ref.~\cite{2023PDU....4201298A} as well as a transition between deceleration and acceleration \cite{2007PhRvD..76d1301M,2024JHEAp..42..178A,2022MNRAS.509.5399C}.}. Evidently, one may notice that at the equivalenve time it must be $G\gtrsim1$, under the hypothesis that the source for $G(z)$ is a generic barotropic fluid.

Immediately, one can compute the cosmographic terms, namely the deceleration $q$, the jerk $j$, and the snap $s$ parameters \cite{2016IJGMM..1330002D,visser}. In
particular, from the definition of the deceleration parameter
\begin{equation}
    \label{qudef}
q(z)=-1+(1+z)\frac{H^\prime(z)}{H(z)}\,,
\end{equation}
having indicated with ${y}^\prime$ the derivative of $y$ with respect to $z$, from Eq.~\eqref{hz} we infer
\begin{subequations}
\begin{align}
\label{qudef2}
q(z)&=\frac{1}{2}+\frac{\Omega_r (1 + z)^4 + \Omega_{DE}[(1+z)G^\prime - 3 G]}{2  [\Omega_m(1 + z)^3 + \Omega_r(1 + z)^4 +  \Omega_{DE} G]}\,,\\
\label{jeidef}
j(z)&=(1+z)q^\prime(z)+2q(z)^2+q(z)\,,\\
\label{essedef}
s(z)&=-(1+z) j^\prime(z) - j(z) \left[2 + 3 q(z)\right]\,.
\end{align}
\end{subequations}

The above quantities are \emph{general}, above all, \emph{model-independent}, and can be calculated at any redshift, representing the so-called \emph{cosmographic set} \cite{2010dmap.conf..287V,2007CQGra..24.5985C}.

\subsection{Physical meaning of  equivalence among cosmological species}

At the epoch of matter domination, one can approximate the cosmic fluid with matter and radiation only. Analogously, at current time, one can approximate the universe as composed by matter and dark energy only.
Consequently, dealing with the equivalence between two fluids implies the existence of a cosmological epoch, namely matter-dominated epoch in the first example, while dark energy-dominated epoch in the second one.

However, since a fluid evolves in time, it is always possible to ensure an equivalence between \emph{any} subdominant fluids.
For example, baryons are always subdominant than cold dark matter, but it is clearly possible to presume that the densities of baryons and radiation turn out to be of same order around,
\begin{equation}\label{esempio}
    z_{br}=-1 + \frac{\Omega_b}{\Omega_{r}}=539^{+18}_{-18},
\end{equation}
where the baryonic density $\Omega_b=0.04930^{+0.00086}_{-0.00086}$ and the radiation density $\Omega_r=9.15^{+0.26}_{-0.26}\times10^{-5}$ have been considered \cite{Planck2018}.
However, in the case of Eq.~\eqref{esempio}, we cannot conclude that a cosmological epoch exists, because at $z_{br}$ the dominant species is the  cold dark matter.

Nevertheless, examining the times when two fluids reach equivalence, even if they remain subdominant compared to matter, could provide hints towards  their evolution.
Thus, the ability to constrain the functional forms of the underlying fluids in a model-independent manner would enable to understand their evolution over time.
Accordingly, if we explore the equivalence times between dark energy and other species, albeit such species may not define distinct cosmological epochs, they can still offer clues about dark energy nature.
To clarify this point, we first focus on the cosmographic equivalence between dark energy and matter and, later, we pass to the equivalence between dark energy and radiation that will occur during the matter domination epoch.

\subsection{The cosmographic dark energy--matter equivalence redshift}

The concept of \emph{equivalence redshift} is frequently associated with the most popular equivalence between radiation and matter, that however occured in the very early Universe \cite{2023PDU....4201298A,Muccino:2020gqt}, approximately at $z\simeq 10^4$.

However, at intermediate times, there is the need to formulate the existence of a further redshift, roughly corresponding to the \emph{equivalence between dark energy and matter} \cite{2023PDU....4201298A}.
This occurs by virtue of the functional form of matter and dark energy. So, one would expect that such an  equivalence time can be used rather as a \emph{cosmographic discriminator} toward the understanding of the right cosmological background, \textit{i.e.}, giving hints on the nature of dark energy  throughout the Universe evolution. Obtaining possible cosmographic bounds on transition times would therefore clarify whether and how much dark energy evolves.

We here go beyond the standard recipe to search for the equivalence between dark energy and matter, by investigating the redshift, $z_{drm}$, at which the dark energy density becomes equivalent to the sum of radiation and matter densities, fulfilling the condition
\begin{equation}\label{formal0}
\Omega_{DE} G_{drm}=\Omega_{r}(1+z_{drm})^4+\Omega_m(1+z_{drm})^3\,,
\end{equation}
where hereafter the subscript $drm$ can be written for a generic function as $X_{drm}\equiv X(z_{drm})$.

In particular, this epoch
\begin{itemize}
\item[-] may occur at a \textit{smaller redshift} than the transition between dark energy and dark matter;
\item[-] differently from the onset of cosmic acceleration, where the deceleration parameter identically vanishes  \cite{2018A&A...616A..32M}, here $q_{drm}\neq0$. This is an advantage since if one expands $q$ around $z_{tr}$, the first order is zero and, so, to fix constraints on $q$ one has to reach at least the second order. Instead, for $z_{drm}$, the first order is not zero and so bounds can be more easily found at first order of Taylor expansion.
\end{itemize}

To fix priors on our model-independent treatment, we can now provide some forecasts on the  equivalence redshifts in the contexts of  three well-know background cosmologies \cite{2006IJMPD..15.1753C,Li:2012dt}:
\begin{itemize}
\item[-] the $\Lambda$CDM paradigm, with $G^\Lambda(z)=1$ at all $z$,
\item[-] the $w$CDM model with $G^w(z)\equiv(1+z)^{3(1+w)}$, and
\item[-] the CPL model \cite{Chevallier2001,Linder2003} with the position $G^{\rm CPL}(z)\equiv(1+z)^{3(1+w_0+w_1)}\exp{\left[-3w_1z/(1+z)\right]}$.
\end{itemize}
Consequently, Eq.~\eqref{formal0} for the $\Lambda$CDM, $w$CDM, and CPL scenarios becomes, respectively
\begin{subequations}
\begin{align}
\label{formal0L}
\Omega_{DE} &= \Omega_r(1+z_{drm})^4 + \Omega_m (1+z_{drm})^3\,,\\
\label{formal0w}
G^w_{drm}\Omega_{DE} &= \Omega_r(1+z_{drm})^4 + \Omega_m (1+z_{drm})^3\,,\\
\label{formal0CPL}
G^{\rm CPL}_{drm}\Omega_{DE} &= \Omega_r(1+z_{drm})^4 + \Omega_m (1+z_{drm})^3\,.
\end{align}
\end{subequations}
Using best-fit values for $\Lambda$CDM, $w$CDM and CPL models \cite{Planck2018}, from Eqs.~\eqref{formal0L}--\eqref{formal0CPL} the equivalence occurs at
\begin{subequations}
\begin{align}
z_{drm}^\Lambda &= 0.295^{+0.015}_{-0.015}\,,\label{priorsmodels1}\\
z_{drm}^w &= 0.310^{+0.050}_{-0.041}\,,\label{priorsmodels2}\\
z_{drm}^{\rm CPL} &= 0.298^{+0.063}_{-0.045}\,.\label{priorsmodels3}
\end{align}
\end{subequations}
where we used for all the models  $\Omega_m=0.3153^{+0.0073}_{-0.0073}$ and $\Omega_{DE}=0.6847^{+0.0073}_{-0.0073}$. The additional parameter of the $w$CDM model is given by $w=-1.028^{+0.031}_{-0.031}$, whereas the additional ones for the CPL parametrization are $w_0=-0.957^{+0.080}_{-0.080}$ and  $w_1=-0.29^{+0.32}_{-0.26}$.

Thus, the redshifts in Eqs. \eqref{priorsmodels1}--\eqref{priorsmodels3} are easily accessible by fitting the data of low-redshift probes, such as SNe Ia \cite{2018ApJ...859..101S}, BAOs \cite{2019JCAP...10..044C} or cosmic chronometers \cite{2023IJMPD..3250039K}.

Clearly, the above expected values are in quite perfect agreement with the equivalence redshifts between matter and dark energy as obtained e.g. in Ref.~\cite{2023PDU....4201298A}, indicating that the radiation density parameter is negligible in obtaining bounds with dark energy, as expected.
Accordingly, in view of these considerations, in Sec.~\ref{sez3drm} the contribution of the radiation $\Omega_r(1+z_{drm})^4$ in the cosmographic reconstruction at the equivalence redshifts $z_{drm}$ will be neglected.

A quite different situation occurs if matter is negligible with respect to radiation. In the case of radiation dominated Universe, assuming that dark energy evolves in the three scenarios indicated before, we will elucidate below which kind of constraints are expected, showing that a possible equivalence between species may occur at much higher redshifts than Eqs.~\eqref{priorsmodels1}--\eqref{priorsmodels3}.

\subsection{The cosmographic dark energy--radiation  equivalence redshift}

As stated above, invoking a cosmographic equivalence between dark energy and radiation would

\begin{itemize}
    \item[-] be strongly dependent on the model under exam,
    \item[-] discriminate, if computed model-independently, how dark energy behaves at intermediate times,
    \item[-] occur at higher redshifts than the dark energy - matter equivalence,
    \item[-] turn out to be less accessible with current data,
    \item[-] hold likewise $q_{eq}\neq0$, as for the case of $z_{drm}$.
\end{itemize}

Thus, equating dark energy and radiation magnitudes implies a quite different situation than above, accordingly giving through Eq.~\eqref{hz},
\begin{equation}\label{formal}
\Omega_{DE}G_{eq}=\Omega_{r}(1+z_{eq})^4\,.
\end{equation}
Resorting again the three cosmological models, above described, namely the $\Lambda$CDM, $w$CDM and CPL scenarios, from  Eq.~\eqref{formal}, we obtain the following constraints
\begin{subequations}
\begin{align}
\label{formal1}
\Omega_{DE}&=\Omega_{r}(1+z_{eq})^4\,,\\
\label{formal1wCDM}
G^w_{eq}\Omega_{DE}&=\Omega_{r}(1+z_{eq})^4\,,\\
\label{formal1CPL}
G^{\rm CPL}_{eq}\Omega_{DE}&=\Omega_{r}(1+z_{eq})^4\,.
\end{align}
\end{subequations}
Eqs.~\eqref{formal1}--\eqref{formal1wCDM} are solved analytically, whereas Eq.~\eqref{formal1CPL} is solved numerically.
Using the already-defined best-fit values for the three scenarios, the equivalence between radiation and dark energy shall occur at
\begin{subequations}
\begin{align}
z_{eq}^\Lambda &= 8.300^{+0.070}_{-0.070}\,,\label{zeqL1}\\
z_{eq}^w &= 7.944^{+0.551}_{-0.497}\,,\label{zeqL2}\\
z_{eq}^{\rm CPL}&=6.746^{+4.123}_{-1.875}\,.\label{zeqL3}
\end{align}
\end{subequations}

Looking at Eqs.~\eqref{zeqL1}--\eqref{zeqL3}, we draw the considerations summarized below.
\begin{itemize}
\item[-] The equivalence in this case is particularly unstable. Different dark energy paradigms, that degenerate at late times, provide quite different priors on the equivalence redshift, even slightly modifying the free parameters.
\item[-] The forecasts span within high-redshift intervals, naively indicating that dark energy scenarios require the use of indicators that cannot be SNe Ia, BAO or OHD data points.
\item[-] High-redshift probes like GRBs are the only astrophysical objects placed at intermediate redshifts and able to estimate such redshifts.
\end{itemize}

In the next section, we shall fix direct bounds on our equivalence redshifts, adopting the aforementioned priors and employing the cosmographic reconstructions, in order to check whether or not the standard cosmological model is predictive enough.

\section{Cosmographic reconstruction of the equivalence redshifts}\label{sec:3}

We here consider the simplest dark energy case satisfying the conditions in Eqs.~\eqref{condizioni}. This implies $G_{drm}>1$ and $G_{eq}>1$, since both $z_{drm}>0$ and $z_{eq}>0$.

We thus single out the cosmographic cases of equivalence among dark energy and matter first, and then dark energy with radiation.

\subsection{Cosmographic bound between  dark energy and matter equivalence}\label{sez3drm}

This first case occurs at late times, where the equivalence between dark energy and matter happens.

From Eq.~\eqref{formal0}, neglecting the radiation and fixing $\Omega_{DE}=1-\Omega_m$, we obtain
\begin{equation}
\label{zdrm1}
z_{drm}= \left[G_{drm}\left(\frac{1-\Omega_m}{\Omega_{m}}\right)\right]^{\frac{1}{3}}-1\,.
\end{equation}
Since $\Omega_r\approx0$ in Eq.~\eqref{hz}, then $\Omega_m$ can be found by inverting Eq.~\eqref{zdrm1}. This procedure is particularly useful as it effectively eliminates the matter density, which typically degenerates with $H_0$ \cite{2002astro.ph..3225R}, thereby bypassing the degeneracy problem altogether.

Plugging the expression for $\Omega_m$ into Eqs.~\eqref{qudef2}--\eqref{essedef}, at $z_{drm}$ we obtain
\begin{subequations}
\label{cosmoparzdrm}
\begin{align}
q_{drm}&=-\frac{1}{4}+ (1+z_{drm})\frac{G_{drm}^\prime}{4G_{drm}},\\
j_{drm}&=1+\frac{f_{drm}}{4G_{drm}},\\
s_{drm}&=-\frac{5}{4}-(1 + z_{drm})\frac{G_{drm}^\prime f_{drm} + G_{drm}g_{drm}}{16G_{drm}^2},
\end{align}
\end{subequations}
with the definitions
\begin{align}
\nonumber
f_{drm}&\equiv (1+z_{drm})\left[G_{drm}^{\prime\prime}(1+z_{drm})-2G_{drm}^\prime\right],\\
\nonumber
g_{drm}&\equiv 6G_{drm}^\prime +(1+z_{drm}) [4G_{drm}^{\prime\prime\prime}(1+z_{drm})- G_{drm}^{\prime\prime}].
\end{align}

We will use the above relations when adopting the GRB data, as we will clarify later.

\subsection{Cosmographic bound between dark energy and radiation equivalence}

Following the same argument reported above, from Eq.~\eqref{formal}, since $\Omega_{DE}=1-\Omega_m-\Omega_r$, it is possible to single out $\Omega_m$.
Afterwards, considering Eq. \eqref{hz} and resorting Eqs.~\eqref{qudef2}--\eqref{essedef}, at the equivalence we obtain
\begin{subequations}
\label{cosmoparzeq}
\begin{align}
q_{eq}=&\, \frac{G_{eq}-(1 + z_{eq})^4\Omega_r-m_{eq}}{g_{eq}}\,,\\
j_{eq}=&\,-\frac{2(f_{eq}+m_{eq})-n_{eq}}{g_{eq}}\,,\\
s_{eq}=&\,-(1 + z_{eq})j^\prime_{eq}-j_{eq}(2+3q_{eq})\,,
\end{align}
\end{subequations}
where we introduced the following definitions
\begin{align}
\nonumber
f_{eq}&\equiv \left[2G_{eq}-(1+z_{eq})^3\right]\Omega_r\,,\\
\nonumber
g_{eq}&\equiv 2 \left[G_{eq}\left(1-\Omega_r\right)+(1+z_{eq})f_{eq} \right]\,,\\
\nonumber
m_{eq}&\equiv \left[G_{eq}-(1+z_{eq})^2 G_{eq}^\prime\right]\Omega_r\,,\\
\nonumber
n_{eq}&\equiv \left[(1+z_{eq})^3(2z_{eq}-G_{eq}^{\prime\prime})-8G_{eq}z_{eq}\right]\Omega_r-2G_{eq}\,.
\end{align}

Again, these relations will be compared with cosmic data and, particularly, with GRBs. We need now to focus on cosmographic reconstructions of distances in order to remove the model-dependence into our computation.

\subsection{Distance reconstructions}

Every Hubble rate can be expanded and the derivatives can be compared directly with data points. So, taking Eq.~\eqref{hz} in series of $\Delta z_x=z-z_x$, we write
\begin{equation}
\label{Hubbleth}
H_{th} = H_x \left[1 + \mathcal H_1^x \Delta z_x + \mathcal H_2^x \Delta z_x^2 + \mathcal H_3^x \Delta z_x^3\right]\,,
\end{equation}
where, depending on the kind of cosmographic reconstruction, ``$x$'' may indicate either ``$drm$'' or ``$eq$''.
At $z=0$, Eq.~\eqref{Hubbleth} constrains to $H_0$, yielding \cite{Orlando}
\begin{equation}
\label{Hubblefinalth}
H_{th} = H_0 \left[\frac{1 + \mathcal H_1^x \Delta z_x + \mathcal H_2^x \Delta z_x^2 + \mathcal H_3^x \Delta z_x^3}{1 + \mathcal H_1^x z_x + \mathcal H_2^x z_x^2 + \mathcal H_3^x z_x^3}\right]\,,
\end{equation}
where we baptize
\begin{subequations}
\begin{align}
\mathcal H_1^x&= \frac{1 + q_x}{1+z_x}\,,\quad \mathcal H_2^x=\frac{j_x-q_x^2}{2(1+z_x)^2}\,,\\
\mathcal H_3^x&=\frac{3 q_x^2 (1 + q_x) - j_x (3 + 4 q_x) - s_x}{6 (1 + z_x)^3}\,.
\end{align}
\end{subequations}
In the above relations, we can utilize the above definitions given in Eqs.~\eqref{cosmoparzdrm}--\eqref{cosmoparzeq}.

Once the Hubble rate in Eq.~\eqref{Hubblefinalth} is defined, to explore cosmological bounds over cosmographic reconstructions at the equivalence times, the luminosity distance (in Mpc units) for a spatially-flat Universe can be derived as
\begin{equation}
\label{dlth}
d_{th}(z) = \left(1+z\right) \int_0^z\dfrac{dz'}{H_{th}(z')}\,,
\end{equation}
providing a simple expression for distance modulus,
\begin{equation}
\label{muz}
\mu_{th}(z)= 25+5\log\left[d_{th}(z)\right]\,.
\end{equation}

The subscript ``${\rm th}$'' stresses that the here-involved quantities are directly predicted from a \emph{theoretical framework} and, so, unless specified differently later on, appear model-dependent, postulating $H_{th}$ within them.

Bearing Eqs.~\eqref{dlth}--\eqref{muz} with the previous theoretical definition in Eqs.~\eqref{cosmoparzdrm}--\eqref{cosmoparzeq}, we now have all the ingredients to employ our cosmic data and perform model-independent analyses as we report below.


\section{Forecasting GRB data}\label{sec:4}

As already mentioned, GRBs are the only astrophysical sources that can probe the Universe at $z_{eq}$ and, possibly, provide useful constraints on the corresponding cosmographic reconstruction \cite{Luongo:2021pjs}.

However, in the last two decades, only two bursts lit-up the $\gamma$-ray sky at such large distances: GRB 090423 with a spectroscopic redshift $z=8.2$ \cite{Tanvir2009} and GRB 090429B with a photometric redshift $z\sim9.4$ \cite{Cucchiara2011}.
Future missions, among all THESEUS \cite{2021ExA....52..183A}, might observe $\approx100$ GRBs at $z\gtrsim5$ over the first $3$ years of activity, with possible detections up to $z\sim14$--$15$ which would shed light into:
\begin{itemize}
    \item[-] Universe reionization era, that according to Planck satellite mission is placed around $z = 7.68 \pm 0.79$,
    \item[-] dark energy--radiation equivalence, placed around $z\simeq 7$--$9$, as predicted from Eqs.~\eqref{zeqL1}--\eqref{zeqL3}.
\end{itemize}

Hence, forecasting GRB data is needful with the purpose of addressing the lack of GRB data points lying around the redshifts where equivalence occurs.

Consequently, to obtain a more extensive data set of GRBs we can produce a set of both simulated and real observational points, covering a range of redshifts that roughly extends up to $z\simeq 12$.

To do so, we make use of the $E_{\rm p}$--$E_{\rm iso}$ correlation, often referred to as the \textit{Amati relation}, \cite{2002A&A...390...81A,AmatiDellaValle2013}
\begin{equation}
\label{Amatirel}
\log E_{\rm p} = a \left(\log E_{\rm iso}-52\right) + b\,.
\end{equation}
More precisely, Eq.~\eqref{Amatirel} represents a linear correlation, characterized by a slope $a$, an intercept $b$ and an extra source of variability $\sigma$, established between the GRB spectral peak energy computed in the source rest-frame $E_{\rm p}$ (in keV units), and the isotropic equivalent energy radiated in $\gamma$-rays (in erg units)
\begin{equation}\label{eiso}
 E_{\rm iso}\equiv 4\pi d_l^2 S_{\rm b}(1+z)^{-1}\,,
\end{equation}
In Eq.~\eqref{eiso} the observed bolometric GRB fluence $S_{\rm b}$ is determined by integrating the energy spectrum in the rest-frame energy range of $1-10^4$ keV, whereas  $d_l$ denotes the luminosity to the GRB source.

In this analysis, we make use of the most up-to-date catalog of $N_A=118$ long GRBs from Ref.~\citep{2021JCAP...09..042K}, characterized by a strong correlation between $E_{\rm p}$ and $E_{\rm iso}$ and the smallest intrinsic dispersion.

\begin{figure*}[!t]
{\hfill
\includegraphics[width=0.45\hsize,clip]{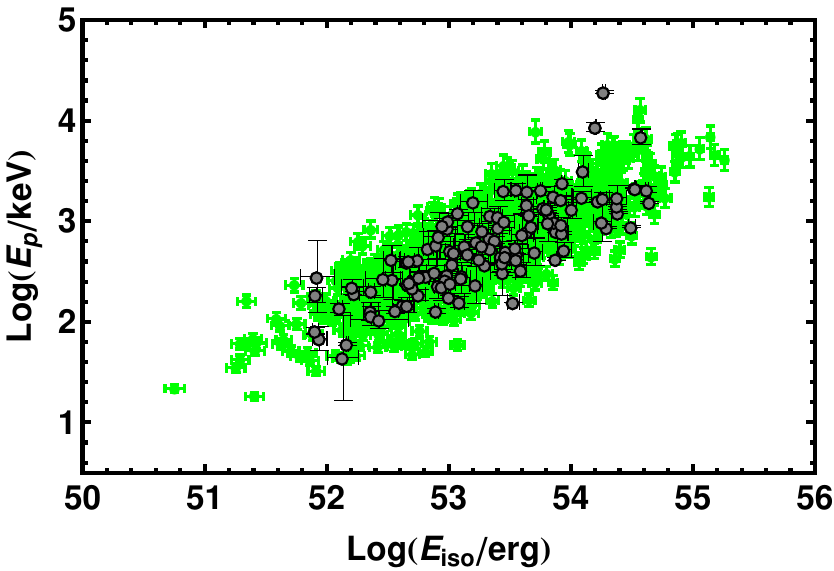}
\hfill
\includegraphics[width=0.47\hsize,clip]
{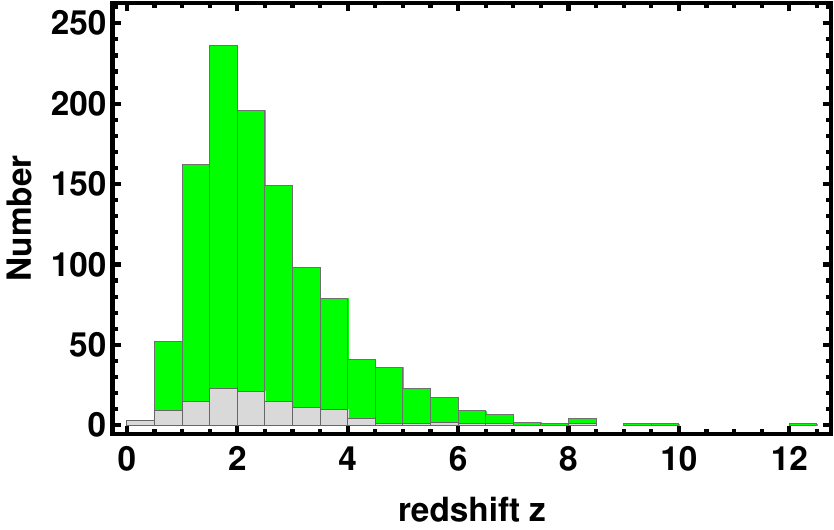}
\hfill}
\caption{\textit{Left}: the comparison between the observed $E_{{\rm p},i}$--$E_{{\rm iso},i}$ sample \citep{2021JCAP...09..042K} (grey data) and the simulated $E_{{\rm p},j}$--$E_{{\rm iso},j}$ one (green data).
\textit{Right}: the comparison between the distributions of the observed (grey chart) and the simulated (green chart) redshifts.}
\label{fig:sim}
\end{figure*}

\subsection{The simulated $E_{\rm p}$--$E_{\rm iso}$ data set}

In addition to the above $N_{\rm A}$ sources, we simulate other $N_{\rm s}=1000$ GRBs, fulfilling the $E_{\rm p}$--$E_{\rm iso}$ correlation.
This is motivated by the fact that, with a total number $N = N_{\rm A} + N_{\rm s} = 1118$, the overall GRB catalog becomes comparable with the \textit{Pantheon} data set of SNe Ia \cite{2018ApJ...859..101S} and with the most updated sample of quasars \cite{2019NatAs...3..272R}.

As a first step, we need to determine the $E_{\rm p}$--$E_{\rm iso}$ correlation parameters ($a,b,\sigma$).
We use the $\Lambda$CDM paradigm best-fit values \cite{Planck2018} to determine the luminosities distances\footnote{All the simulated $\log E_{{\rm iso},j}$ points are generated using the $\Lambda$CDM model. This introduces neither circularity, nor model-dependence issues, as we will show, first introducing the model-independent calibration via B\'ezier polynomials, and then recalling that $G(z)\rightarrow1$ as $z\rightarrow0$, justifying the $\Lambda$CDM model use.} $d_{l}(z_i)$ of the observed $N_A$ sources and to compute their isotropic energies $E_{\rm iso}(z_i)$ from Eq.~\eqref{eiso}.

The correlation parameters can be determined by maximizing the log-likelihood
\begin{equation}
\label{a1}
\ln \mathcal{L}_{\rm A} = -\sum_{i=1}^{N_{\rm A}}\left\{\dfrac{\left[Y_i-Y(z_i)\right]^2}{2\sigma_{ Y_i}^2} + \ln(\sqrt{2\pi}\sigma_{Y_i})\right\},\\
\end{equation}
where we  defined,
\begin{subequations}
\begin{align}
Y_{\rm i} \equiv&\, \log E_{{\rm p},i}\,,\\
\label{a2}
Y(z_i)\equiv&\, a \left[\log E_{\rm iso} (z_i)-52\right] + b\,,\\
\sigma_{Y_i}^2 \equiv&\, \sigma_{\log E_{{\rm p},i}}^2 + a^2\sigma_{\log E_{{\rm iso},i}}^2+\sigma^2\,.
\end{align}
\end{subequations}
The corresponding MCMC best-fit parameters are
\begin{subequations}
\label{parsLCDM}
\begin{align}
a=&\,0.537_{-0.039}^{+0.036}\,,\\ b=&\,2.043_{-0.055}^{+0.057}\,,\\
\sigma=&\,0.257_{-0.018}^{+0.018}\,.
\end{align}
\end{subequations}
The simulated data can be built up from the $N_{\rm A}$ observed GRBs by following the recipe, reported below \cite{2007MNRAS.379L..55L}.
\begin{itemize}
\item[-] The observed redshifts $\log z_i$ obey a normal distribution characterized by a mean value $\mu_z=0.359$ and a variance  $\sigma_z=0.214$.
\item[-] From the above normal distribution with $\mu_z$ and $\sigma_z$, we generate $N_{\rm s}$ redshifts $\log z_j$.
\item[-] The observed isotropic energies $\log E_{{\rm iso},i}$, in units of $10^{52}$~erg, follow a normal distribution with a mean value $\mu_E=1.300$ and a variance $\sigma_z=0.718$.
\item[-] From the above log-normal distribution with $\mu_E$ and $\sigma_E$, we generate $N_{\rm s}$ isotropic energies $\log E_{{\rm iso},j}$.
\item[-] For each pair ($\log z_j$, $\log E_{{\rm iso},j}$) we generate the peak energy $\log E_{{\rm p},j}$ from the normal distribution of mean value $\mu_p=a(\log E_{{\rm iso},j}-52) + b$ and variance $\sigma$, where ($a,b,\sigma$) are taken from Eqs.~\eqref{parsLCDM}.
\item[-] The simulated pairs ($\log E_{{\rm p},j},\log E_{{\rm iso},j}$) naturally satisfy the $E_{\rm p}$--$E_{\rm iso}$ correlation, once the best-fit parameters, listed in Eqs.~\eqref{parsLCDM}, are considered.
\item[-] The simulated errors on $\log E_{{\rm p},j}$ are generated by computing the mean error of the observed peak energies $\langle \sigma_{\log E_{{\rm p},i}}\rangle$ and weighting each of them with the ratio of the simulated peak energy and the mean of the observed peak energies $\langle \log E_{{\rm p},i}\rangle$, namely
\begin{equation}
\label{errorEpsim}
\sigma_{\log E_{{\rm p},j}} = \langle \sigma_{\log E_{{\rm p},i}}\rangle \frac{\log E_{{\rm p},j}}{\langle \log E_{{\rm p},i}\rangle}\,.
\end{equation}
\item[-] Similarly, the simulated errors on $\log E_{{\rm iso},j}$ are generated by computing the mean error of the observed isotropic energies $\langle \sigma_{\log E_{{\rm iso},i}}\rangle$ and the mean of the observed isotropic energies $\langle \log E_{{\rm iso},i}\rangle$, namely
\begin{equation}
\label{errorEisosim}
\sigma_{\log E_{{\rm iso},j}} = \langle \sigma_{\log E_{{\rm iso},i}}\rangle \frac{\log E_{{\rm iso},j}}{\langle \log E_{{\rm iso},i}\rangle}\,.
\end{equation}
\end{itemize}

In this respect, Fig.~\ref{fig:sim} displays the comparison between the observed data set and the simulated catalog (gray versus green points, respectively).
In particular, the right plot portrays the redshift distributions of observed and simulated data, evidencing that our initial goal of expanding the data set of GRBs around the likely value of $z_{eq}$ has been achieved with additional $10$ additional sources within the range, $7\lesssim z\lesssim12$.

\subsection{Calibration of the $E_{\rm p}$--$E_{\rm iso}$ correlation}

As already noticed from Eq.~\eqref{eiso}, the isotropic energy measurement is affected by the well-known \emph{circularity problem} \cite{Kodama2008}, meaning that its determination depends upon an \emph{a priori imposition of a background cosmology} due to  the luminosity distance $d_l$.

This issue is overcome only by calibrating $E_{\rm iso}$ by means of model-independent techniques.
Here, we resort a well-established strategy, as shown in Refs. ~\cite{2019MNRAS.486L..46A,LM2020,2021MNRAS.501.3515M,2023MNRAS.518.2247L,2023MNRAS.523.4938M}, based on the interpolation of the $N_{\rm O}=32$ Hubble rate measurements  \cite{2023IJMPD..3250039K}, provided by cosmic chronometers \cite{2002ApJ...573...37J}, by using a second order B\'ezier parametric curve,
\begin{equation}
\label{bezier1}
H_2(x) = 100 \left(\frac{\rm km/s}{\rm Mpc}\right) \sum_{i=0}^{2} 2\alpha_i \frac{x^i}{i!}\frac{\left(1-x\right)^{2-i}}{(2-i)!}\,,
\end{equation}
where $\alpha_i$ are the coefficients of the linear combination and the variable $x\equiv z/z_{\rm m}$ depends upon the maximum redshift $z_{\rm m}=1.965$ of the observational Hubble data, obtained from the use of cosmic chronometers.

To estimate the coefficients $\alpha_i$ from cosmic chronometers, we maximize the log-likelihood function,
\begin{equation}
\label{loglikeOHD}
    \ln \mathcal{L}_{\rm O} = -\sum_{k=1}^{N_{\rm O}}\left\{\dfrac{\left[H_k-H_2(z_k)\right]^2}{2\sigma_{H_k}^2} + \ln(\sqrt{2\pi}\sigma_{H_k})\right\}\,,
\end{equation}
where the errors $\sigma_{H_k}$ include the statistical uncertainties \cite{2023IJMPD..3250039K} and a careful evaluation of the systematic uncertainties \cite{2020ApJ...898...82M,2022LRR....25....6M,2023MNRAS.523.4938M}.
The MCMC best-fit values are
\begin{subequations}
\label{alphaOHD}
\begin{align}
\alpha_0 &= 0.651^{+0.091}_{-0.092}\,,\\
\alpha_1 &= 1.111^{+0.219}_{-0.223}\,,\\
\alpha_2 &= 2.042^{+0.248}_{-0.246}\,,
\end{align}
\end{subequations}
in strict agreement with the results found in Ref.~\cite{2023MNRAS.523.4938M}.

Bearing in mind the assumption $\Omega_k=0$ and the results of Eqs.~\eqref{alphaOHD}, we can obtain a cosmology-independent evaluation of the luminosity distance
\begin{equation}
\label{dlHz2}
d_2(z)=\left(1+z\right)\int_0^z\dfrac{dz'}{H_2(z')}\,,
\end{equation}
enabling us to obtain a calibrated isotropic energy
\begin{equation}
\label{Eisocal}
E_2(z)\equiv 4\pi d_2^2(z) S_{\rm b}(1+z)^{-1}\,,
\end{equation}
where the respective errors on $E_2(z)$ depend on those related to both $S_{\rm b}$ and $H_2(z)$, propagating to $d_2(z)$.

\section{Cosmological bounds on the equivalence redshifts}
\label{sec:5}

Following the  guidelines provided above, and using the entire set of observed and simulated GRBs (totaling N), we can determine the correlation parameters ($a$, $b$, $\sigma$) by assessing a calibration log-likelihood. Simultaneously, we can establish the cosmological parameters at the dark energy -- radiation and matter equivalence ($h_0$, $z_{drm}$, $G_{drm}$, $G_{drm}^\prime$, $G_{drm}^{\prime\prime}$, $G_{drm}^{\prime\prime\prime}$) or at the dark energy -- radiation equivalence ($h_0$, $z_{eq}$, $\Omega_r$, $G_{eq}$, $G_{eq}^\prime$, $G_{eq}^{\prime\prime}$, $G_{eq}^{\prime\prime\prime}$) through a cosmological log-likelihood.

As stated in the introduction, to accomplish this, we employ two methods, conventionally named methods A and B, as described below.
\begin{itemize}
\item[-] {\bf Method A} involves the independent maximization of the calibration and cosmological log-likelihood functions using the complete set of N GRBs. Initially, the calibration log-likelihood determines the correlation parameters along with their associated uncertainties. Subsequently, these parameters are utilized in the cosmological log-likelihood to assess the cosmological parameters.
\item[-] {\bf Method B}, essentially, is a hierarchical Bayesian regression (HBR). The strategy combines two log-likelihood functions, the first encompassing a calibrator sample of GRBs with redshifts falling within the observational range of cosmic chronometers ($z \leq z_m$), whereas the second consisting of a cosmological sample, comprising the entire GRB dataset.
\end{itemize}

For the numerical analyses, the subsequent priors on the correlation parameters are assumed,
\begin{equation}
\nonumber
a\in \left[0,2\right],\quad\quad b\in \left[0,3\right],\quad\quad \sigma\in \left[0,1\right],
\end{equation}
and, analogously, for the cosmographic parameters,
\begin{equation}
\nonumber
\begin{array}{rclcrclr}
h_0 & \in & \left[0,1\right] & , & \Omega_r & \in &\left[0,0.0004\right] &,\\
z_{drm} & \in & \left[0,2\right] & , &
z_{eq} & \in & \left[0,15\right] &,\\
G_{drm}\ {\rm,}\ G_{eq} & \in & \left[0,3\right] & , & G^\prime_{drm}\ {\rm,}\ G^\prime_{eq} & \in & \left[-10,10\right] &,\\
G^{\prime\prime}_{drm}\ {\rm,}\ G^{\prime\prime}_{eq} & \in & \left[-10,10\right] & , & G^{\prime\prime\prime}_{drm}\ {\rm,}\ G^{\prime\prime\prime}_{eq} & \in & \left[-10,10\right] &.
\end{array}
\end{equation}

Our findings are thus split for methods A and B.

\begin{table*}
\centering
\setlength{\tabcolsep}{0.2em}
\renewcommand{\arraystretch}{1.4}
\begin{tabular}{lcccccccccc}
\hline\hline
                &  $a$
                &  $b$
                &  $\sigma$
                &  $h_0$
                &  $z_{drm}$
                &  --
                &  $G_{drm}$
                &  $G_{drm}^\prime$
                &  $G_{drm}^{\prime\prime}$
                &  $G_{drm}^{\prime\prime\prime}$\\
\hline
A               & $0.537_{-0.019}^{+0.017}$
                & $2.036_{-0.023}^{+0.029}$
                & $0.191_{-0.011}^{+0.012}$
                & $0.846_{-0.182}^{+0.149}$
                & $0.709_{-0.420}^{+0.379}$
                & --
                & $2.678_{-0.912}^{+0.313}$
                & $-0.395_{-2.284}^{+1.333}$
                & $-1.896_{-2.792}^{+4.191}$
                & $0.278_{-3.280}^{+3.428}$\\
B               & $0.708_{-0.015}^{+0.018}$
                & $1.848_{-0.030}^{+0.023}$
                & $0.199_{-0.011}^{+0.012}$
                & $0.701_{-0.122}^{+0.176}$
                & $0.518_{-0.288}^{+0.563}$
                & --
                & $2.709_{-1.081}^{+0.291}$
                & $0.530_{-1.592}^{+1.234}$
                & $-3.694_{-1.190}^{+4.547}$
                & $-1.176_{-3.361}^{+4.081}$\\
\hline
                &  $a$
                &  $b$
                &  $\sigma$
                &  $h_0$
                &  $z_{eq}$
                &  $10^4\times\Omega_r$
                &  $G_{eq}$
                &  $G_{eq}^\prime$
                &  $G_{eq}^{\prime\prime}$
                &  $G_{eq}^{\prime\prime\prime}$\\
\hline
A               &  $0.537_{-0.019}^{+0.017}$
                &  $2.036_{-0.023}^{+0.029}$
                &  $0.191_{-0.011}^{+0.012}$
                &  $0.825_{-0.117}^{+0.079}$
                &  $5.583_{-2.093}^{+3.572}$
                &  $2.471_{-1.424}^{+1.477}$
                &  $0.968_{-0.896}^{+1.328}$
                &  $-2.711_{-2.080}^{+6.046}$
                &  $3.173_{-3.840}^{+1.824}$
                &  $-1.869_{-2.836}^{+5.685}$\\
B               &  $0.727_{-0.022}^{+0.017}$
                &  $1.788_{-0.028}^{+0.037}$
                &  $0.210_{-0.011}^{+0.014}$
                &  $0.712_{-0.096}^{+0.086}$
                &  $7.223_{-3.264}^{+1.010}$
                &  $1.746_{-0.886}^{+1.812}$
                &  $1.628_{-1.083}^{+1.094}$
                &  $3.060_{-6.502}^{+1.617}$
                &  $3.822_{-8.363}^{+1.161}$
                &  $-2.820_{-1.945}^{+6.883}$\\
\hline
\end{tabular}
\caption{Methods A and B best-fit parameters and $1$--$\sigma$ errors obtained from the whole sample of $N$ GRBs for the cosmographic approaches around dark energy--radiation (top part of Table) and dark energy--matter (bottom part of Table) equivalences.}
\label{tab:summarymodelslike}
\end{table*}

\subsection{Numerical constraints: method A}

Within this approach, the calibration log-likelihood is the same as in Eq.~\eqref{a1} but now it runs over $N$ sources.

In addition, the isotropic energies of the whole sample are now calibrated through the B\'ezier interpolation, as reported in Eq.~\eqref{Eisocal}, and the corresponding errors also account for the uncertainties on $H_2(z)$.
Thus, we have
\begin{subequations}
\label{Ycalibration}
\begin{align}
Y(z_i)\equiv&\, a \left[\log E_2(z_i)-52\right] + b\,,\\
\sigma_{Y_i}^2 \equiv&\, \sigma_{\log E_{{\rm p},i}}^2 + a^2\left(\sigma_{\log S_{{\rm b},i}}^2 + 4\sigma_{\log d_{2,i}}^2\right) + \sigma^2\,.
\end{align}
\end{subequations}
The MCMC fit for the calibration log-likelihood is the same for both the cosmographic approaches around the dark energy--radiation equivalence and around the dark energy--radiation and matter equivalence.
The results are listed in Table~\ref{tab:summarymodelslike} and portrayed in Fig.~\ref{fig:simA}.

The cosmological log-likelihood is given by
\begin{equation}
\label{a4}
\ln \mathcal{L}_{\rm C} = -\sum_{j=1}^{N}\left\{\dfrac{\left[\mu_j-\mu_{\rm th}(z_j)\right]^2}{2\sigma_{\mu_j}^2} + \ln (\sqrt{2\pi} \sigma_{\mu_j})\right\}\,,
\end{equation}
where we have resorted the GRB distance moduli  obtained from the $E_{\rm p}$--$E_{\rm iso}$ correlation and the corresponding errors, respectively \cite{LM2020,2023MNRAS.518.2247L,2023MNRAS.523.4938M}
\begin{subequations}
\begin{align}
\mu_j\equiv& \,\frac{5}{2 a}\left[\log E_{{\rm p},j} - a\log\left(\frac{4\pi S_{{\rm b},j}}{1+z_j}\right) - b\right]\,,\\
\nonumber
\sigma_{\mu_j}^2 \equiv& \,\frac{25}{4 a^2}\left(\sigma_{\log E_{{\rm p},j}}^2 + a^2 \sigma_{\log S_{{\rm b},j}}^2+\sigma^2\right)+\left(\frac{\partial \mu_j}{\partial a}\right)^2\sigma_a^2 +\\
\label{a5}
&\,2\frac{\partial \mu_j}{\partial a}\frac{\partial \mu_j}{\partial b} \sigma_{ab} + \left(\frac{\partial \mu_j}{\partial b}\right)^2\sigma_b^2\,,
\end{align}
\end{subequations}
where $\sigma_{a}$, $\sigma_b$ and $\sigma_{ab}$ are the covariance terms between the parameters $a$ and $b$, needed since the calibration and the cosmological MCMC fits are separately performed.

The results of the MCMC fit for the cosmological log-likelihood, for both cosmographic approaches, are listed in Table~\ref{tab:summarymodelslike} and portrayed in Fig.~\ref{fig:simA}.

\subsection{Numerical constraints: method B}

The extrapolation of $H_2(z)$ at $z>z_{\rm m}$ may bias the calibration of the Amati correlation and, thus, the evaluation of the cosmological parameters \cite{LM2020}.
To reduce this possible issue, the HBR combines two nested samples,
\begin{itemize}
\item[1)] the calibrator sub-sample of GRBs, composed of $N_{\rm cal}=685$ sources with redshifts $z\leq z_{\rm m}$, used to estimate the correlation parameters, and
\item[2)] the cosmological sample, with all the $N$ GRBs, used to estimate the cosmological parameters.
\end{itemize}

Thus, the total log-likelihood function is given by
\begin{equation}
    \ln \mathcal{L} = \ln \mathcal{L}_{\rm A} + \ln \mathcal{L}_{\rm C}\,.
\end{equation}

For the log-likelihood $\ln \mathcal L_{\rm A}$, the definitions in Eqs.~\eqref{a1} and \eqref{Ycalibration} still hold, with the prescription that the sum runs over the $N_{\rm cal}$ GRBs of the calibrator sub-sample.

The log-likelihood $\ln \mathcal L_{\rm C}$ is the same as in Eq.~\eqref{a4} with the only difference of the distance modulus errors
\begin{equation}
\sigma_{\mu_j}^2 \equiv \,\frac{25}{4 a^2}\left(\sigma_{\log E_{{\rm p},j}}^2 + a^2 \sigma_{\log S_{{\rm b},j}}^2+\sigma^2\right)\,,
\end{equation}
where the covariance terms $\sigma_{a}$, $\sigma_b$ and $\sigma_{ab}$ are no longer needed, since the HBR approach computes $a$, $b$ and $\sigma$ together with the other parameters and provides an unique covariance matrix for all the parameters.

The results of the MCMC fits related to the Method B, for both cosmographic approaches around the radiation--dark energy equivalence
and around the radiation and matter--dark energy equivalence, are listed in Table~\ref{tab:summarymodelslike} and displayed in Figs.~\ref{fig:simBdrm} and \ref{fig:simB}, respectively.

\section{Theoretical implications of our findings}\label{sec:6}

We here focus on theoretical interpretation and the physical meaning of the two equivalence redshifts, that appear quite different as emphasized in Secs.~\ref{intro} and \ref{sec:2}.

We  remark the main differences between inferring a cosmological epoch (where a species dominates) and a cosmographic time discriminator (where two subdominant species equate). Afterwards, the bounds for $z_{drm}$ and $z_{eq}$, respectively at small and large redshifts, are reported and commented. In this respect, we argue how dark energy may evolve throughout the entire Universe evolution, up to the last simulated GRB redshift bin.

\subsection{Interpreting \emph{cosmographic time discriminator}}

When we equate the magnitudes of matter, radiation, and dark energy, we can get evidence for the existence of distinct cosmological epochs, characterized by varying combinations of the three mentioned components.

As stated in Sec.~\ref{sec:2}, a cosmological epoch arises when a given constituent dominates over the other cosmic fluids at a given time.
Examples are the equivalences between matter and radiation and/or dark energy with matter.
The underlying approximation is valid when we consider an Einstein-de Sitter Universe, in which one fluid only drives the overall cosmological dynamics.

Conversely, computing the equivalence between dark energy and radiation magnitudes would imply the same as above only if there exists an epoch where radiation dominates over dark energy and a subsequent time where dark energy dominates over radiation. However, the presence of matter complicates this scenario and, in fact, at redshift $z\simeq7$--$10$, as predicted by Eqs.~\eqref{zeqL1}--\eqref{zeqL3}, radiation is significantly less dominant than matter.

Consequently, dark energy and radiation magnitudes can be the same, but matter would dominate over them and, in turn, we cannot deal with a proper cosmological epoch, but rather with a cosmographic time that discriminates how and whether dark energy is dynamical. Phrasing it differently, we seek \emph{a cosmographic time discriminator} associated with the form of dark energy.

Clearly, this process is identified under certain circumstances, such as,
\begin{itemize}
\item[-] there is no interaction among constituents, i.e., no interactions between dark energy and radiation, or between dark energy and matter and so forth occur,
\item[-] the fluids evolve with precise equations of state,
\item[-] dark energy does not increase dramatically as the Universe expands,
\item[-] no early dark energy has been postulated,
\item[-] spatial curvature is negligible throughout the Universe's history and does not influence the analysis.
\end{itemize}

The last condition deserves a better clarification. While, spatial curvature does not have a substantial impact on the equivalence at low redshifts, around $z\lesssim1$, its influence may become more pronounced as $z$ increases.

All these theoretical considerations justify the focus on our two equivalences: one at very low redshifts, involving dark energy with matter plus radiation, and the subsequent equivalence between dark energy and radiation.

\subsection{Analyzing the results}

{\bf Method A.} The results of the MCMC fits are summarized in Table~\ref{tab:summarymodelslike} and portrayed in Fig.~\ref{fig:simA}.
We recall that method A performs calibration and cosmological fits separately, thus neglecting the hierarchy of the log-likelihood functions and the mutual influence of both correlation and cosmological parameters.
This is particularly evident looking at the bounds on the cosmographic parameters.
\begin{itemize}
    \item[-] {\bf Equivalence at $z_{drm}$}. With respect to the $\Lambda$CDM predictions, the cosmographic parameters $h_0$, $z_{drm}$ and $G_{drm}$ exhibit quite high values. However, if on the one hand $h_0$ and $z_{drm}$ are consistent with the concordance model due to the huge attached errors, on the other $G_{drm}$ is incompatible with the expectation $G(z)=1$, though its derivatives within the huge errors are consistent with the cosmological constant conjecture.
    \item[-] {\bf Equivalence at $z_{eq}$}. In this epoch, $h_0$ and $z_{eq}$ are, respectively, higher and lower than (but within the errors consistent with) the concordance model values. The resulting $\Omega_r$ parameter is higher and barely inconsistent with the value provided by the Planck satellite \cite{Planck2018}. Conversely, $G_{eq}$ and its derivatives are consistent -- within huge errors -- with the standard cosmological constant scenario.
\end{itemize}

{\bf Method B.} The results of the MCMC are listed in Table~\ref{tab:summarymodelslike} and portrayed in Figs.~\ref{fig:simBdrm} and \ref{fig:simB}.
\begin{itemize}
    \item[-] {\bf Equivalence at $z_{drm}$}. The values of $h_0$ and $z_{drm}$ are still higher but closer to those predicted by the $\Lambda$CDM model. Further,  $G_{drm}$ is here incompatible with the concordance model.
    \item[-] {\bf Equivalence at $z_{eq}$}. In this case, within  large errors, all the parameters are consistent with the cosmological concordance background.
\end{itemize}

From the results of the two methods, we draw the conclusions listed below.
\begin{itemize}
    \item[-] The Bayesian approach provides improved results with respect to the method A, in particular on $z_{eq}$.
    \item[-]  The epoch $z_{drm}$ provides the most controversial constraint in both methods, i.e., $G_{drm}>1$. This may be seen as a direct consequence of the shortage of low-redshift GRBs. Even including the $N_s$ simulated GRBs, the number of sources at $z\lesssim 0.6$ is just $10$, too scarce to obtain acceptable and reliable constraints for the equivalence at low redshifts.
    \item[-] The sample of GRBs is large enough to provide constraints that are in line with the expectations of the concordance paradigm.
\end{itemize}

In view of these outcomes, we notice that contrasting conclusions may arise in view of the \emph{dark energy evolution} and the \emph{Hubble constant tension}.
\begin{itemize}
    \item[-] {\bf Equivalence at $z_{drm}$}. For both the methods, the results suggest $G_{drm}>1$ with null derivatives. Both sets of constraints are consistent with the dark energy under the form of a constant or of a slowly evolving function of time (nearly constant). On the other hand, the large attached errors do not provide enough elements for a definitive conclusion.
    \item[-] {\bf Equivalence at $z_{eq}$}. Conversely to $z_{drm}$, $G_{eq}$ and its derivatives favor the cosmological constant scenario. However, the large errors cannot firmly exclude the possibility that dark energy may slowly evolve with time, as well as for the $z_{drm}$ case.
\end{itemize}

In addition, it is worth noticing what follows.
\begin{itemize}
    \item[-] The results of the equivalence epoch at $z_{drm}$ are jeopardized by the lack of sources at low redshift, preventing us to obtain reliable estimates.
    \item[-] Though not conclusive, the above considerations tend to favor the cosmological constant scenario purported by the $\Lambda$CDM paradigm.
    \item[-] The Hubble constant tension is not fully-addressed, since all values of $h_0$ in Table~\ref{tab:summarymodelslike} are consistent, within the errors, with both Planck estimate in the flat scenario \cite{Planck2018} and $h_0=0.7304\pm0.0104$ inferred from Cepheids \citep{2022ApJ...934L...7R}. It is worth noticing that Method B results are in line with a recent estimate got from SNe Ia based on surface brightness fluctuations measurements, i.e., $h_0=0.7050\pm0.0237$ \citep{2021A&A...647A..72K}, that seems to indicate that the Hubble constant may be in between the extreme values currently in mutual tension. So, even though our methods are model-independent, the existing tension is far from reaching its solution even adopting GRBs.
\end{itemize}

Finally, we compare our findings with those of Ref.~\cite{2023PDU....4201298A} and summarized the implications of the current work.
\begin{itemize}
    \item[-] We went further in the cosmographic reconstruction of dark energy equation of state, up to $G^{\prime\prime\prime}(z)$.
    \item[-] We investigated such a reconstruction also for the high-redshift equivalence epoch at $z_{eq}$.
    \item[-] We utilized only GRBs to provide constraints on the two equivalence epochs.
    \item[-] We did not include the spatial curvature in our analysis for mainly two reasons: first the calibration of GRB data with the interpolation of the Hubble rate data is not straightforward when the spatial curvature is accounted for, and, second, in Ref.~\cite{2023PDU....4201298A}, it was shown that its inclusion did not significantly change the results.
\end{itemize}

\section{Final outlooks and perspectives}\label{sec:7}

In this work we investigated the cosmic reconstructions of two different equivalence epochs between dark energy and a) radiation and b) radiation and matter.

We discussed the physical meaning of these two epochs. Then, we distinguished the equivalence as a cosmological epoch, where one of the two fluids dominates over all the others, from the equivalence between two subdominant fluids. In the latter, we introduced the concept of cosmographic time discriminator, showing that this equivalence may serve as discriminator among cosmological models.

To work this procedure out, we selected  GRB data from the well-consolidated $E_{\rm p}$--$E_{\rm iso}$ correlation.
To overcome the \emph{circularity problem} \cite{Kodama2008}, GRB data were calibrated through the well-established B\'ezier interpolation of the Hubble rate measurements \cite{2019MNRAS.486L..46A,LM2020,2021MNRAS.501.3515M,2023MNRAS.518.2247L,2023MNRAS.523.4938M}.
GRBs represent unique astrophysical sources that lap $z_{eq}$ and future missions, such as  THESEUS \cite{2021ExA....52..183A}, will provide wider numbers of GRB detections at $z\gtrsim5$, i.e., considerably increasing the catalog and possibly helping in shedding light around the equivalence epochs here considered. In this respect, for consistency, we used GRBs alone also for getting bounds on the low-redshift equivalence at $z_{drm}$.

However, to increase the statistic and the quality of our constraints, we simulated $1000$ additional GRBs fulfilling the $E_{\rm p}$--$E_{\rm iso}$ correlation so that the redshift coverage spans up to $z\simeq12$. In particular, we showed that if an equivalence between subdominant species like dark energy and radiation occurs, then it happens in the matter-dominated era, i.e., at $z\gtrsim7$, being strongly dependent on the free parameters of a given dark energy model.

In this respect, we considered two kind of MCMC fitting methods to extract cosmic bounds: in the first, GRB calibration and cosmological fits are performed independently, whereas in the second, essentially a hierarchical Bayesian regression, are performed simultaneously.

The outcomes of our analysis showed that:
\begin{itemize}
    \item[-] Method B provided better results than method A. However, both outcomes appeared more in line with the expectations of the concordance paradigm, in particular for the $z_{eq}$ bounds.
    \item[-] The value $G_{drm}>1$ implied a larger value of $z_{drm}$, though consistent with the $\Lambda$CDM scenario, within the error bars. This issue is very likely due to shortage of GRBs (observed and simulated) at low redshifts that affects the bounds got from the cosmographic reconstruction during this epoch.
    \item[-] Besides the above issue resulting in $G_{drm}>1$ at $z_{drm}$, both methods indicated that dark energy is mainly in the form of a cosmological constant or slightly  evolving with time. Precisely, the bounds from the equivalence at $z_{eq}$ are more in line with the concordance model expectations though. However, the large attached errors did not provide enough elements for a definitive conclusion.
    \item[-] The Hubble constant tension was not fully-fixed, albeit the more trustworthy results, got from the equivalence at $z_{eq}$, provided bounds that seemed to indicate that the Hubble constant may lie between the well-known two values measured in Refs.~\cite{Planck2018,2022ApJ...934L...7R}.
\end{itemize}

Summarizing, our outcomes, though non-conclusive, confirmed the general trend that the $\Lambda$CDM model is the most suited one to describe the cosmic evolution, at least with the current level of precision of GRB data.

Certainly, as perspectives, new and more precise data from GRB surveys can improve the situation, especially for accessing the high-redshift domains. Indeed, our simulated GRB data reflect the current level of accuracy for GRB observables and include also the additional errors (both statistical and systematic) of the Hubble rate measurements, introduced by the calibration procedure.

Nevertheless, to compensate the lack of low-redshift GRBs and improve cosmographic reconstructions around $z_{drm}$, it is desirable to include also low-redshift catalogs such as SNe Ia, Hubble rate measurements, BAOs and so on, and find the most suited technique to combine them.

As a final hint, we will use our methods to check more complicated dark energy scenarios, even involving plausible interactions among dark constituents and/or adding the spatial curvature as well. Further, the results provided by the DESI mission \cite{2024arXiv240403002D} will be included to refine the analysis and to check whether departures on the equivalence times will be expected.

\section*{Acknowledgements}
The authors are grateful to Kuantay Boshkayev and Peter K.~S.~Dunsby for very useful discussions on the topic of cosmological models. OL is thankful to Alejandro Aviles, Francesco Pace and Sunny Vagnozzi for intriguing debates on the nature of dark energy and on the impact of the last cosmological findings toward a possible evolution of dark energy. MM acknowledges Lorenzo Amati, Massimo Della Valle, Luca Izzo and Luca Porcelli for valuable comments on the numerical analyses.

\appendix

\section{Contour plots from Methods A and B}\label{appA}

We here show the contour plots obtained from MCMC analyses of Methods A and B for both the cosmographic approaches around the dark energy--radiation and dark energy--radiation and matter equivalences.

\begin{figure*}[!t]
\includegraphics[width=0.59\hsize,clip]{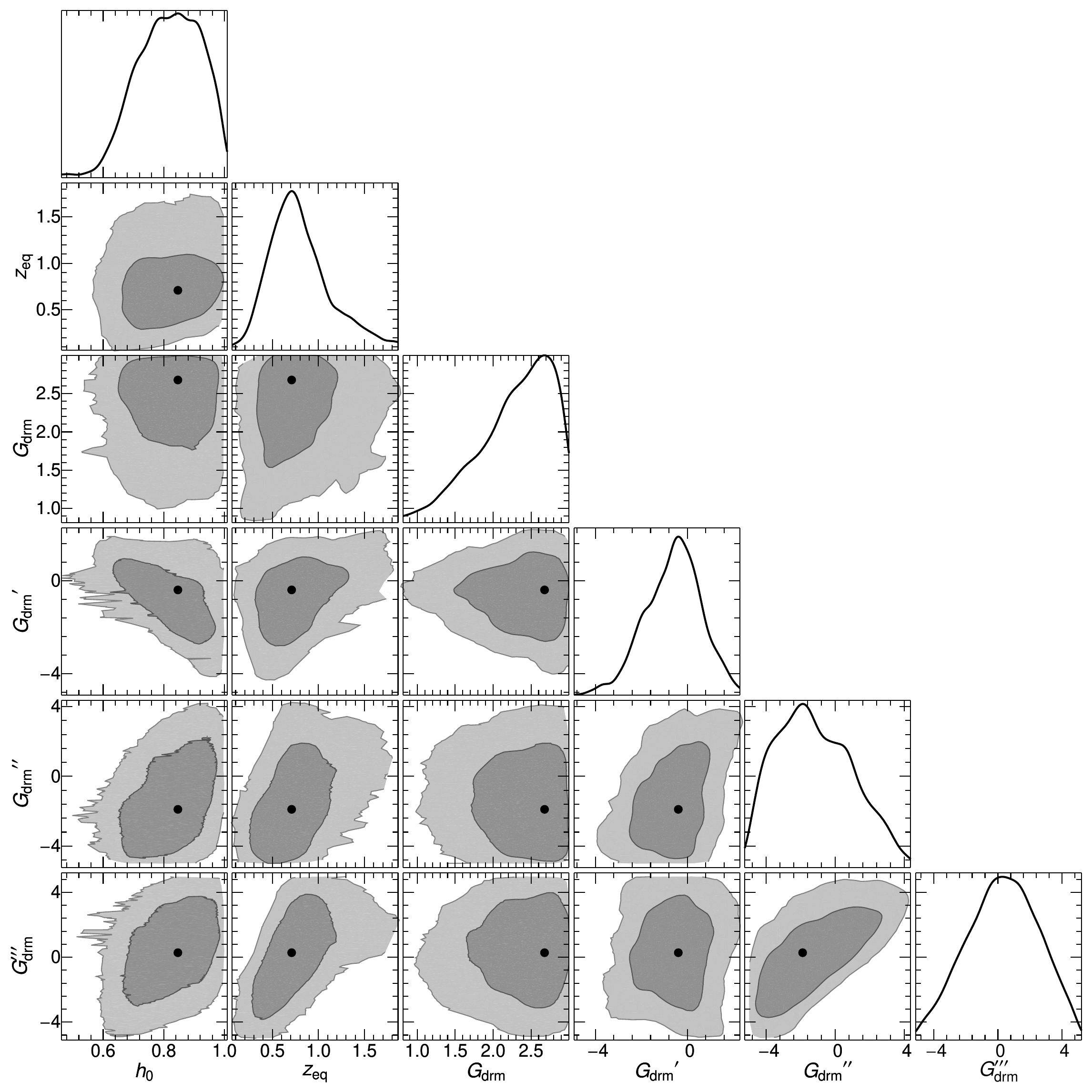}\hfill
\includegraphics[width=0.31\hsize,clip]{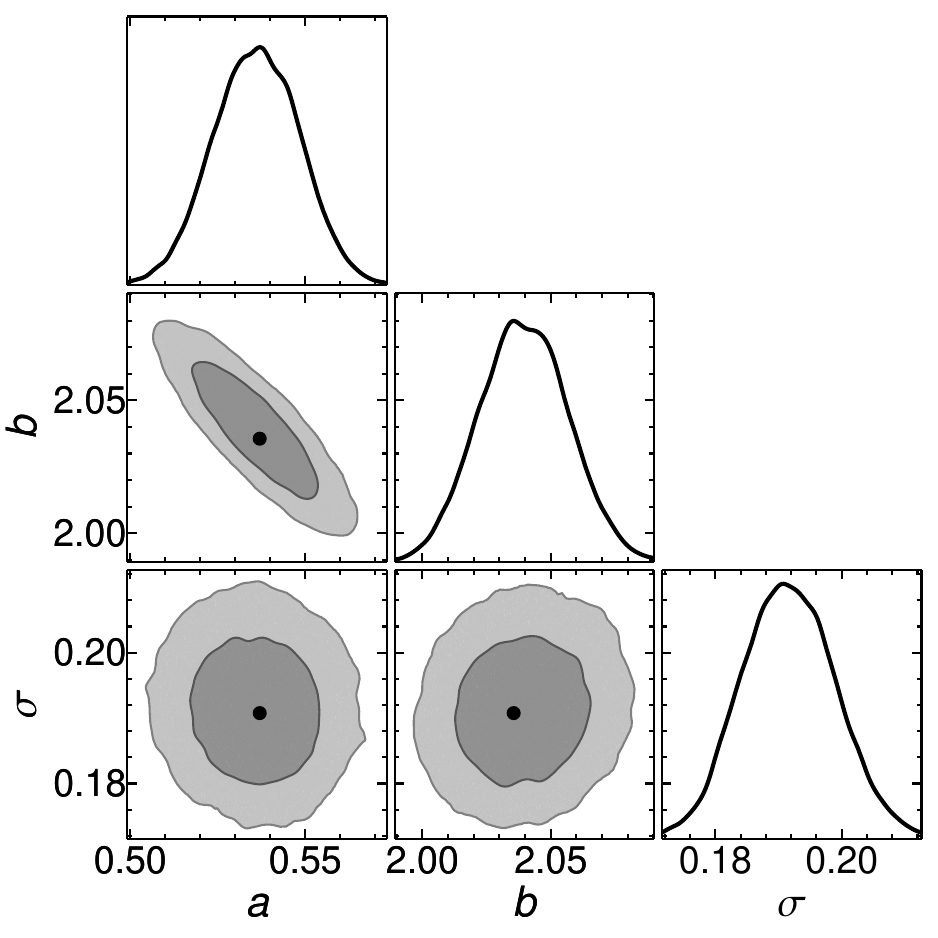}
\includegraphics[width=0.68\hsize,clip]{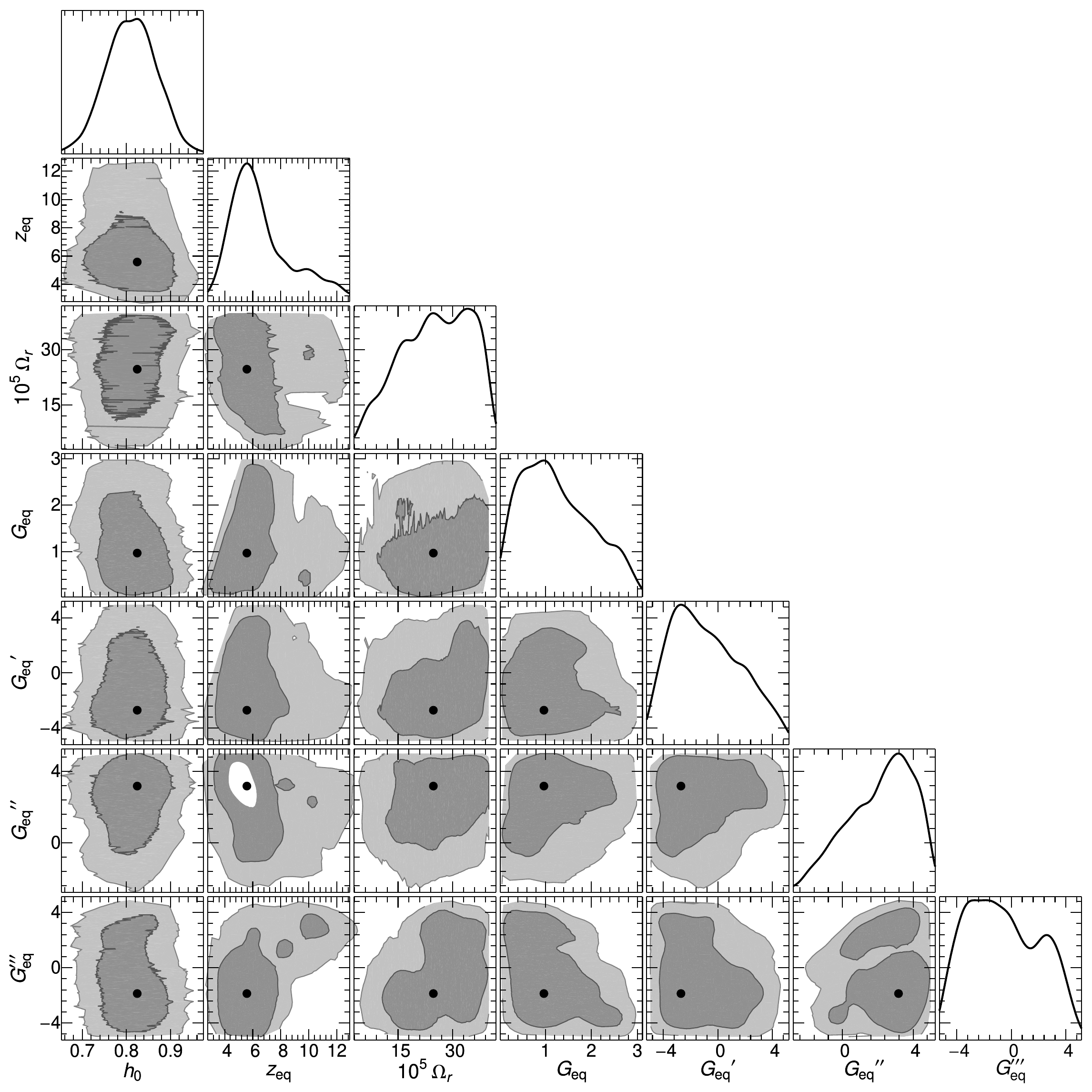}\hfill
\includegraphics[width=0.31\hsize,clip]{MCMC_GRB_corr_real32.pdf}
\caption{Method A contour plots from the calibrated $E_{\rm p}$--$E_{\rm iso}$ correlation. \textit{Top panels}: the best-fit parameters at the dark energy -- matter and radiation equivalence. \textit{Bottom panels}: the best-fit parameters at the dark energy -- radiation equivalence. Darker (lighter) areas mark the $1$--$\sigma$ ($2$--$\sigma$) confidence regions.}
\label{fig:simA}
\end{figure*}

\begin{figure*}[!t]
\includegraphics[width=0.9\hsize,clip]{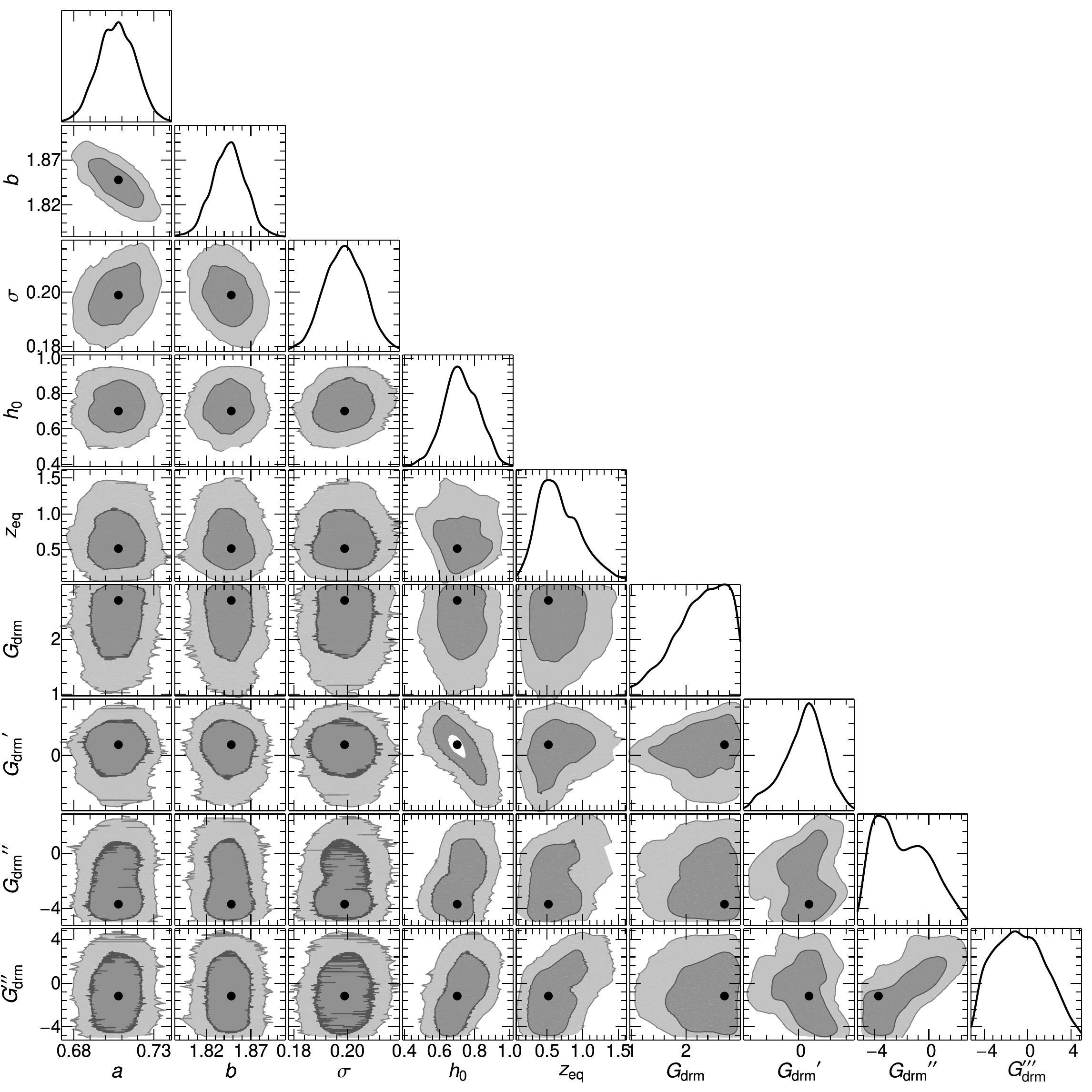}
\caption{Method B contour plots, got from the HBR technique, of the best-fit parameters of the $E_{\rm p}$--$E_{\rm iso}$ correlation and the cosmographic approach at the dark energy -- matter and radiation equivalence. Darker (lighter) areas mark the $1$--$\sigma$ ($2$--$\sigma$) confidence regions.}
\label{fig:simBdrm}
\end{figure*}

\begin{figure*}[!t]
\includegraphics[width=\hsize,clip]{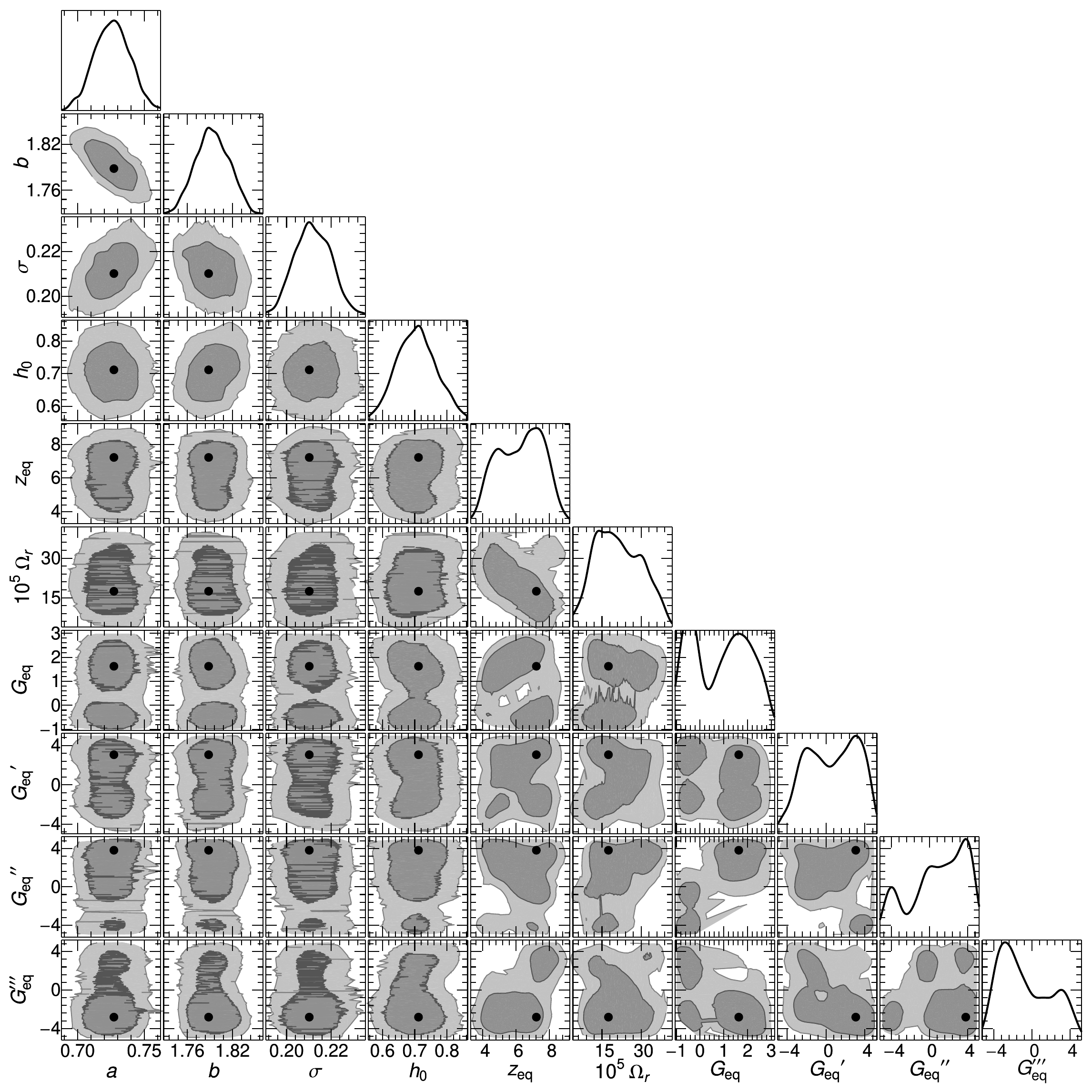}
\caption{Method B contour plots, got from the HBR technique, of the best-fit parameters of the $E_{\rm p}$--$E_{\rm iso}$ correlation and the cosmographic approach at the dark energy -- radiation equivalence. Darker (lighter) areas mark the $1$--$\sigma$ ($2$--$\sigma$) confidence regions.}
\label{fig:simB}
\end{figure*}


\begin{thebibliography}{93}
\expandafter\ifx\csname natexlab\endcsname\relax\def\natexlab#1{#1}\fi
\expandafter\ifx\csname bibnamefont\endcsname\relax
  \def\bibnamefont#1{#1}\fi
\expandafter\ifx\csname bibfnamefont\endcsname\relax
  \def\bibfnamefont#1{#1}\fi
\expandafter\ifx\csname citenamefont\endcsname\relax
  \def\citenamefont#1{#1}\fi
\expandafter\ifx\csname url\endcsname\relax
  \def\url#1{\texttt{#1}}\fi
\expandafter\ifx\csname urlprefix\endcsname\relax\def\urlprefix{URL }\fi
\providecommand{\bibinfo}[2]{#2}
\providecommand{\eprint}[2][]{\url{#2}}

\bibitem[{\citenamefont{{Planck Collaboration}}(2020)}]{Planck2018}
\bibinfo{author}{\bibnamefont{{Planck Collaboration}}}, \bibinfo{journal}{A\&A}
  \textbf{\bibinfo{volume}{641}}, \bibinfo{eid}{A6} (\bibinfo{year}{2020}),
  \eprint{1807.06209}.

\bibitem[{\citenamefont{{Perlmutter} et~al.}(1998)\citenamefont{{Perlmutter},
  {Aldering}, {della Valle}, {Deustua}, {Ellis}, {Fabbro}, {Fruchter},
  {Goldhaber}, {Groom}, {Hook} et~al.}}]{1998Natur.391...51P}
\bibinfo{author}{\bibfnamefont{S.}~\bibnamefont{{Perlmutter}}},
  \bibinfo{author}{\bibfnamefont{G.}~\bibnamefont{{Aldering}}},
  \bibinfo{author}{\bibfnamefont{M.}~\bibnamefont{{della Valle}}},
  \bibinfo{author}{\bibfnamefont{S.}~\bibnamefont{{Deustua}}},
  \bibinfo{author}{\bibfnamefont{R.~S.} \bibnamefont{{Ellis}}},
  \bibinfo{author}{\bibfnamefont{S.}~\bibnamefont{{Fabbro}}},
  \bibinfo{author}{\bibfnamefont{A.}~\bibnamefont{{Fruchter}}},
  \bibinfo{author}{\bibfnamefont{G.}~\bibnamefont{{Goldhaber}}},
  \bibinfo{author}{\bibfnamefont{D.~E.} \bibnamefont{{Groom}}},
  \bibinfo{author}{\bibfnamefont{I.~M.} \bibnamefont{{Hook}}},
  \bibnamefont{et~al.}, \bibinfo{journal}{\nat} \textbf{\bibinfo{volume}{391}},
  \bibinfo{pages}{51} (\bibinfo{year}{1998}), \eprint{astro-ph/9712212}.

\bibitem[{\citenamefont{{Riess} et~al.}(1998)\citenamefont{{Riess},
  {Filippenko}, {Challis}, {Clocchiatti}, {Diercks}, {Garnavich}, {Gilliland},
  {Hogan}, {Jha}, {Kirshner} et~al.}}]{1998AJ....116.1009R}
\bibinfo{author}{\bibfnamefont{A.~G.} \bibnamefont{{Riess}}},
  \bibinfo{author}{\bibfnamefont{A.~V.} \bibnamefont{{Filippenko}}},
  \bibinfo{author}{\bibfnamefont{P.}~\bibnamefont{{Challis}}},
  \bibinfo{author}{\bibfnamefont{A.}~\bibnamefont{{Clocchiatti}}},
  \bibinfo{author}{\bibfnamefont{A.}~\bibnamefont{{Diercks}}},
  \bibinfo{author}{\bibfnamefont{P.~M.} \bibnamefont{{Garnavich}}},
  \bibinfo{author}{\bibfnamefont{R.~L.} \bibnamefont{{Gilliland}}},
  \bibinfo{author}{\bibfnamefont{C.~J.} \bibnamefont{{Hogan}}},
  \bibinfo{author}{\bibfnamefont{S.}~\bibnamefont{{Jha}}},
  \bibinfo{author}{\bibfnamefont{R.~P.} \bibnamefont{{Kirshner}}},
  \bibnamefont{et~al.}, \bibinfo{journal}{\aj} \textbf{\bibinfo{volume}{116}},
  \bibinfo{pages}{1009} (\bibinfo{year}{1998}), \eprint{astro-ph/9805201}.

\bibitem[{\citenamefont{{Perlmutter} et~al.}(1999)\citenamefont{{Perlmutter},
  {Aldering}, {Goldhaber}, {Knop}, {Nugent}, {Castro}, {Deustua}, {Fabbro},
  {Goobar}, {Groom} et~al.}}]{1999ApJ...517..565P}
\bibinfo{author}{\bibfnamefont{S.}~\bibnamefont{{Perlmutter}}},
  \bibinfo{author}{\bibfnamefont{G.}~\bibnamefont{{Aldering}}},
  \bibinfo{author}{\bibfnamefont{G.}~\bibnamefont{{Goldhaber}}},
  \bibinfo{author}{\bibfnamefont{R.~A.} \bibnamefont{{Knop}}},
  \bibinfo{author}{\bibfnamefont{P.}~\bibnamefont{{Nugent}}},
  \bibinfo{author}{\bibfnamefont{P.~G.} \bibnamefont{{Castro}}},
  \bibinfo{author}{\bibfnamefont{S.}~\bibnamefont{{Deustua}}},
  \bibinfo{author}{\bibfnamefont{S.}~\bibnamefont{{Fabbro}}},
  \bibinfo{author}{\bibfnamefont{A.}~\bibnamefont{{Goobar}}},
  \bibinfo{author}{\bibfnamefont{D.~E.} \bibnamefont{{Groom}}},
  \bibnamefont{et~al.}, \bibinfo{journal}{\apj} \textbf{\bibinfo{volume}{517}},
  \bibinfo{pages}{565} (\bibinfo{year}{1999}), \eprint{astro-ph/9812133}.

\bibitem[{\citenamefont{{Tonry} et~al.}(2003)\citenamefont{{Tonry}, {Schmidt},
  {Barris}, {Cand ia}, {Challis}, {Clocchiatti}, {Coil}, {Filippenko},
  {Garnavich}, {Hogan} et~al.}}]{2003ApJ...594....1T}
\bibinfo{author}{\bibfnamefont{J.~L.} \bibnamefont{{Tonry}}},
  \bibinfo{author}{\bibfnamefont{B.~P.} \bibnamefont{{Schmidt}}},
  \bibinfo{author}{\bibfnamefont{B.}~\bibnamefont{{Barris}}},
  \bibinfo{author}{\bibfnamefont{P.}~\bibnamefont{{Cand ia}}},
  \bibinfo{author}{\bibfnamefont{P.}~\bibnamefont{{Challis}}},
  \bibinfo{author}{\bibfnamefont{A.}~\bibnamefont{{Clocchiatti}}},
  \bibinfo{author}{\bibfnamefont{A.~L.} \bibnamefont{{Coil}}},
  \bibinfo{author}{\bibfnamefont{A.~V.} \bibnamefont{{Filippenko}}},
  \bibinfo{author}{\bibfnamefont{P.}~\bibnamefont{{Garnavich}}},
  \bibinfo{author}{\bibfnamefont{C.}~\bibnamefont{{Hogan}}},
  \bibnamefont{et~al.}, \bibinfo{journal}{\apj} \textbf{\bibinfo{volume}{594}},
  \bibinfo{pages}{1} (\bibinfo{year}{2003}), \eprint{astro-ph/0305008}.

\bibitem[{\citenamefont{{Bridle} et~al.}(2003)\citenamefont{{Bridle}, {Lahav},
  {Ostriker}, and {Steinhardt}}}]{2003Sci...299.1532B}
\bibinfo{author}{\bibfnamefont{S.~L.} \bibnamefont{{Bridle}}},
  \bibinfo{author}{\bibfnamefont{O.}~\bibnamefont{{Lahav}}},
  \bibinfo{author}{\bibfnamefont{J.~P.} \bibnamefont{{Ostriker}}},
  \bibnamefont{and} \bibinfo{author}{\bibfnamefont{P.~J.}
  \bibnamefont{{Steinhardt}}}, \bibinfo{journal}{Science}
  \textbf{\bibinfo{volume}{299}}, \bibinfo{pages}{1532} (\bibinfo{year}{2003}),
  \eprint{astro-ph/0303180}.

\bibitem[{\citenamefont{{Bennett} et~al.}(2003)\citenamefont{{Bennett},
  {Halpern}, {Hinshaw}, {Jarosik}, {Kogut}, {Limon}, {Meyer}, {Page},
  {Spergel}, {Tucker} et~al.}}]{2003ApJS..148....1B}
\bibinfo{author}{\bibfnamefont{C.~L.} \bibnamefont{{Bennett}}},
  \bibinfo{author}{\bibfnamefont{M.}~\bibnamefont{{Halpern}}},
  \bibinfo{author}{\bibfnamefont{G.}~\bibnamefont{{Hinshaw}}},
  \bibinfo{author}{\bibfnamefont{N.}~\bibnamefont{{Jarosik}}},
  \bibinfo{author}{\bibfnamefont{A.}~\bibnamefont{{Kogut}}},
  \bibinfo{author}{\bibfnamefont{M.}~\bibnamefont{{Limon}}},
  \bibinfo{author}{\bibfnamefont{S.~S.} \bibnamefont{{Meyer}}},
  \bibinfo{author}{\bibfnamefont{L.}~\bibnamefont{{Page}}},
  \bibinfo{author}{\bibfnamefont{D.~N.} \bibnamefont{{Spergel}}},
  \bibinfo{author}{\bibfnamefont{G.~S.} \bibnamefont{{Tucker}}},
  \bibnamefont{et~al.}, \bibinfo{journal}{\apjs}
  \textbf{\bibinfo{volume}{148}}, \bibinfo{pages}{1} (\bibinfo{year}{2003}),
  \eprint{astro-ph/0302207}.

\bibitem[{\citenamefont{{Hinshaw} et~al.}(2003)\citenamefont{{Hinshaw},
  {Spergel}, {Verde}, {Hill}, {Meyer}, {Barnes}, {Bennett}, {Halpern},
  {Jarosik}, {Kogut} et~al.}}]{2003ApJS..148..135H}
\bibinfo{author}{\bibfnamefont{G.}~\bibnamefont{{Hinshaw}}},
  \bibinfo{author}{\bibfnamefont{D.~N.} \bibnamefont{{Spergel}}},
  \bibinfo{author}{\bibfnamefont{L.}~\bibnamefont{{Verde}}},
  \bibinfo{author}{\bibfnamefont{R.~S.} \bibnamefont{{Hill}}},
  \bibinfo{author}{\bibfnamefont{S.~S.} \bibnamefont{{Meyer}}},
  \bibinfo{author}{\bibfnamefont{C.}~\bibnamefont{{Barnes}}},
  \bibinfo{author}{\bibfnamefont{C.~L.} \bibnamefont{{Bennett}}},
  \bibinfo{author}{\bibfnamefont{M.}~\bibnamefont{{Halpern}}},
  \bibinfo{author}{\bibfnamefont{N.}~\bibnamefont{{Jarosik}}},
  \bibinfo{author}{\bibfnamefont{A.}~\bibnamefont{{Kogut}}},
  \bibnamefont{et~al.}, \bibinfo{journal}{\apjs}
  \textbf{\bibinfo{volume}{148}}, \bibinfo{pages}{135} (\bibinfo{year}{2003}),
  \eprint{astro-ph/0302217}.

\bibitem[{\citenamefont{{Kogut} et~al.}(2003)\citenamefont{{Kogut}, {Spergel},
  {Barnes}, {Bennett}, {Halpern}, {Hinshaw}, {Jarosik}, {Limon}, {Meyer},
  {Page} et~al.}}]{2003ApJS..148..161K}
\bibinfo{author}{\bibfnamefont{A.}~\bibnamefont{{Kogut}}},
  \bibinfo{author}{\bibfnamefont{D.~N.} \bibnamefont{{Spergel}}},
  \bibinfo{author}{\bibfnamefont{C.}~\bibnamefont{{Barnes}}},
  \bibinfo{author}{\bibfnamefont{C.~L.} \bibnamefont{{Bennett}}},
  \bibinfo{author}{\bibfnamefont{M.}~\bibnamefont{{Halpern}}},
  \bibinfo{author}{\bibfnamefont{G.}~\bibnamefont{{Hinshaw}}},
  \bibinfo{author}{\bibfnamefont{N.}~\bibnamefont{{Jarosik}}},
  \bibinfo{author}{\bibfnamefont{M.}~\bibnamefont{{Limon}}},
  \bibinfo{author}{\bibfnamefont{S.~S.} \bibnamefont{{Meyer}}},
  \bibinfo{author}{\bibfnamefont{L.}~\bibnamefont{{Page}}},
  \bibnamefont{et~al.}, \bibinfo{journal}{\apjs}
  \textbf{\bibinfo{volume}{148}}, \bibinfo{pages}{161} (\bibinfo{year}{2003}),
  \eprint{astro-ph/0302213}.

\bibitem[{\citenamefont{{Spergel} et~al.}(2003)\citenamefont{{Spergel},
  {Verde}, {Peiris}, {Komatsu}, {Nolta}, {Bennett}, {Halpern}, {Hinshaw},
  {Jarosik}, {Kogut} et~al.}}]{2003ApJS..148..175S}
\bibinfo{author}{\bibfnamefont{D.~N.} \bibnamefont{{Spergel}}},
  \bibinfo{author}{\bibfnamefont{L.}~\bibnamefont{{Verde}}},
  \bibinfo{author}{\bibfnamefont{H.~V.} \bibnamefont{{Peiris}}},
  \bibinfo{author}{\bibfnamefont{E.}~\bibnamefont{{Komatsu}}},
  \bibinfo{author}{\bibfnamefont{M.~R.} \bibnamefont{{Nolta}}},
  \bibinfo{author}{\bibfnamefont{C.~L.} \bibnamefont{{Bennett}}},
  \bibinfo{author}{\bibfnamefont{M.}~\bibnamefont{{Halpern}}},
  \bibinfo{author}{\bibfnamefont{G.}~\bibnamefont{{Hinshaw}}},
  \bibinfo{author}{\bibfnamefont{N.}~\bibnamefont{{Jarosik}}},
  \bibinfo{author}{\bibfnamefont{A.}~\bibnamefont{{Kogut}}},
  \bibnamefont{et~al.}, \bibinfo{journal}{\apjs}
  \textbf{\bibinfo{volume}{148}}, \bibinfo{pages}{175} (\bibinfo{year}{2003}),
  \eprint{astro-ph/0302209}.

\bibitem[{\citenamefont{{Eisenstein} et~al.}(2005)\citenamefont{{Eisenstein},
  {Zehavi}, {Hogg}, {Scoccimarro}, {Blanton}, {Nichol}, {Scranton}, {Seo},
  {Tegmark}, {Zheng} et~al.}}]{2005ApJ...633..560E}
\bibinfo{author}{\bibfnamefont{D.~J.} \bibnamefont{{Eisenstein}}},
  \bibinfo{author}{\bibfnamefont{I.}~\bibnamefont{{Zehavi}}},
  \bibinfo{author}{\bibfnamefont{D.~W.} \bibnamefont{{Hogg}}},
  \bibinfo{author}{\bibfnamefont{R.}~\bibnamefont{{Scoccimarro}}},
  \bibinfo{author}{\bibfnamefont{M.~R.} \bibnamefont{{Blanton}}},
  \bibinfo{author}{\bibfnamefont{R.~C.} \bibnamefont{{Nichol}}},
  \bibinfo{author}{\bibfnamefont{R.}~\bibnamefont{{Scranton}}},
  \bibinfo{author}{\bibfnamefont{H.-J.} \bibnamefont{{Seo}}},
  \bibinfo{author}{\bibfnamefont{M.}~\bibnamefont{{Tegmark}}},
  \bibinfo{author}{\bibfnamefont{Z.}~\bibnamefont{{Zheng}}},
  \bibnamefont{et~al.}, \bibinfo{journal}{\apj} \textbf{\bibinfo{volume}{633}},
  \bibinfo{pages}{560} (\bibinfo{year}{2005}), \eprint{astro-ph/0501171}.

\bibitem[{\citenamefont{Luongo and Quevedo}(2014)}]{Luongo:2014qoa}
\bibinfo{author}{\bibfnamefont{O.}~\bibnamefont{Luongo}} \bibnamefont{and}
  \bibinfo{author}{\bibfnamefont{H.}~\bibnamefont{Quevedo}},
  \bibinfo{journal}{Phys. Rev. D} \textbf{\bibinfo{volume}{90}},
  \bibinfo{pages}{084032} (\bibinfo{year}{2014}), \eprint{1407.1530}.

\bibitem[{\citenamefont{Bergstrom}(2009)}]{Bergstrom:2009ib}
\bibinfo{author}{\bibfnamefont{L.}~\bibnamefont{Bergstrom}},
  \bibinfo{journal}{New J. Phys.} \textbf{\bibinfo{volume}{11}},
  \bibinfo{pages}{105006} (\bibinfo{year}{2009}), \eprint{0903.4849}.

\bibitem[{\citenamefont{Profumo et~al.}(2019)\citenamefont{Profumo, Giani, and
  Piattella}}]{Profumo:2019ujg}
\bibinfo{author}{\bibfnamefont{S.}~\bibnamefont{Profumo}},
  \bibinfo{author}{\bibfnamefont{L.}~\bibnamefont{Giani}}, \bibnamefont{and}
  \bibinfo{author}{\bibfnamefont{O.~F.} \bibnamefont{Piattella}},
  \bibinfo{journal}{Universe} \textbf{\bibinfo{volume}{5}},
  \bibinfo{pages}{213} (\bibinfo{year}{2019}), \eprint{1910.05610}.

\bibitem[{\citenamefont{{Chevallier} and {Polarski}}(2001)}]{Chevallier2001}
\bibinfo{author}{\bibfnamefont{M.}~\bibnamefont{{Chevallier}}}
  \bibnamefont{and}
  \bibinfo{author}{\bibfnamefont{D.}~\bibnamefont{{Polarski}}},
  \bibinfo{journal}{International Journal of Modern Physics D}
  \textbf{\bibinfo{volume}{10}}, \bibinfo{pages}{213} (\bibinfo{year}{2001}),
  \eprint{gr-qc/0009008}.

\bibitem[{\citenamefont{{Linder}}(2003)}]{Linder2003}
\bibinfo{author}{\bibfnamefont{E.~V.} \bibnamefont{{Linder}}},
  \bibinfo{journal}{Physical Review Letters} \textbf{\bibinfo{volume}{90}},
  \bibinfo{eid}{091301} (\bibinfo{year}{2003}), \eprint{astro-ph/0208512}.

\bibitem[{\citenamefont{{Peebles} and {Ratra}}(2003)}]{2003RvMP...75..559P}
\bibinfo{author}{\bibfnamefont{P.~J.} \bibnamefont{{Peebles}}}
  \bibnamefont{and} \bibinfo{author}{\bibfnamefont{B.}~\bibnamefont{{Ratra}}},
  \bibinfo{journal}{Reviews of Modern Physics} \textbf{\bibinfo{volume}{75}},
  \bibinfo{pages}{559} (\bibinfo{year}{2003}), \eprint{astro-ph/0207347}.

\bibitem[{\citenamefont{{King} et~al.}(2014)\citenamefont{{King}, {Davis},
  {Denney}, {Vestergaard}, and {Watson}}}]{King2014}
\bibinfo{author}{\bibfnamefont{A.~L.} \bibnamefont{{King}}},
  \bibinfo{author}{\bibfnamefont{T.~M.} \bibnamefont{{Davis}}},
  \bibinfo{author}{\bibfnamefont{K.~D.} \bibnamefont{{Denney}}},
  \bibinfo{author}{\bibfnamefont{M.}~\bibnamefont{{Vestergaard}}},
  \bibnamefont{and} \bibinfo{author}{\bibfnamefont{D.}~\bibnamefont{{Watson}}},
  \bibinfo{journal}{\mnras} \textbf{\bibinfo{volume}{441}},
  \bibinfo{pages}{3454} (\bibinfo{year}{2014}), \eprint{1311.2356}.

\bibitem[{\citenamefont{{Padmanabhan}}(2003)}]{2003PhR...380..235P}
\bibinfo{author}{\bibfnamefont{T.}~\bibnamefont{{Padmanabhan}}},
  \bibinfo{journal}{\physrep} \textbf{\bibinfo{volume}{380}},
  \bibinfo{pages}{235} (\bibinfo{year}{2003}), \eprint{hep-th/0212290}.

\bibitem[{\citenamefont{{Sahni} and {Starobinsky}}(2000)}]{2000IJMPD...9..373S}
\bibinfo{author}{\bibfnamefont{V.}~\bibnamefont{{Sahni}}} \bibnamefont{and}
  \bibinfo{author}{\bibfnamefont{A.}~\bibnamefont{{Starobinsky}}},
  \bibinfo{journal}{International Journal of Modern Physics D}
  \textbf{\bibinfo{volume}{9}}, \bibinfo{pages}{373} (\bibinfo{year}{2000}),
  \eprint{astro-ph/9904398}.

\bibitem[{\citenamefont{{Copeland} et~al.}(2006)\citenamefont{{Copeland},
  {Sami}, and {Tsujikawa}}}]{2006IJMPD..15.1753C}
\bibinfo{author}{\bibfnamefont{E.~J.} \bibnamefont{{Copeland}}},
  \bibinfo{author}{\bibfnamefont{M.}~\bibnamefont{{Sami}}}, \bibnamefont{and}
  \bibinfo{author}{\bibfnamefont{S.}~\bibnamefont{{Tsujikawa}}},
  \bibinfo{journal}{International Journal of Modern Physics D}
  \textbf{\bibinfo{volume}{15}}, \bibinfo{pages}{1753} (\bibinfo{year}{2006}),
  \eprint{hep-th/0603057}.

\bibitem[{\citenamefont{{Tsujikawa}}(2011)}]{tsujikawa2011dark}
\bibinfo{author}{\bibfnamefont{S.}~\bibnamefont{{Tsujikawa}}}, in
  \emph{\bibinfo{booktitle}{Astrophysics and Space Science Library}}, edited by
  \bibinfo{editor}{\bibfnamefont{S.}~\bibnamefont{{Matarrese}}},
  \bibinfo{editor}{\bibfnamefont{M.}~\bibnamefont{{Colpi}}},
  \bibinfo{editor}{\bibfnamefont{V.}~\bibnamefont{{Gorini}}}, \bibnamefont{and}
  \bibinfo{editor}{\bibfnamefont{U.}~\bibnamefont{{Moschella}}}
  (\bibinfo{year}{2011}), vol. \bibinfo{volume}{370} of
  \emph{\bibinfo{series}{Astrophysics and Space Science Library}}, p.
  \bibinfo{pages}{331}, \eprint{1004.1493}.

\bibitem[{\citenamefont{{Perivolaropoulos} and
  {Skara}}(2021)}]{2021arXiv210505208P}
\bibinfo{author}{\bibfnamefont{L.}~\bibnamefont{{Perivolaropoulos}}}
  \bibnamefont{and} \bibinfo{author}{\bibfnamefont{F.}~\bibnamefont{{Skara}}},
  \bibinfo{journal}{arXiv e-prints} \bibinfo{eid}{arXiv:2105.05208}
  (\bibinfo{year}{2021}), \eprint{2105.05208}.

\bibitem[{\citenamefont{{Ooba} et~al.}(2018)\citenamefont{{Ooba}, {Ratra}, and
  {Sugiyama}}}]{2018ApJ...864...80O}
\bibinfo{author}{\bibfnamefont{J.}~\bibnamefont{{Ooba}}},
  \bibinfo{author}{\bibfnamefont{B.}~\bibnamefont{{Ratra}}}, \bibnamefont{and}
  \bibinfo{author}{\bibfnamefont{N.}~\bibnamefont{{Sugiyama}}},
  \bibinfo{journal}{\apj} \textbf{\bibinfo{volume}{864}}, \bibinfo{eid}{80}
  (\bibinfo{year}{2018}), \eprint{1707.03452}.

\bibitem[{\citenamefont{{Efstathiou} and
  {Gratton}}(2020)}]{2020MNRAS.496L..91E}
\bibinfo{author}{\bibfnamefont{G.}~\bibnamefont{{Efstathiou}}}
  \bibnamefont{and}
  \bibinfo{author}{\bibfnamefont{S.}~\bibnamefont{{Gratton}}},
  \bibinfo{journal}{\mnras} \textbf{\bibinfo{volume}{496}},
  \bibinfo{pages}{L91} (\bibinfo{year}{2020}), \eprint{2002.06892}.

\bibitem[{\citenamefont{{Luongo} and {Muccino}}(2018{\natexlab{a}})}]{nostro}
\bibinfo{author}{\bibfnamefont{O.}~\bibnamefont{{Luongo}}} \bibnamefont{and}
  \bibinfo{author}{\bibfnamefont{M.}~\bibnamefont{{Muccino}}},
  \bibinfo{journal}{\prd} \textbf{\bibinfo{volume}{98}}, \bibinfo{eid}{103520}
  (\bibinfo{year}{2018}{\natexlab{a}}), \eprint{1807.00180}.

\bibitem[{\citenamefont{{D'Agostino} et~al.}(2022)\citenamefont{{D'Agostino},
  {Luongo}, and {Muccino}}}]{2022CQGra..39s5014D}
\bibinfo{author}{\bibfnamefont{R.}~\bibnamefont{{D'Agostino}}},
  \bibinfo{author}{\bibfnamefont{O.}~\bibnamefont{{Luongo}}}, \bibnamefont{and}
  \bibinfo{author}{\bibfnamefont{M.}~\bibnamefont{{Muccino}}},
  \bibinfo{journal}{Classical and Quantum Gravity}
  \textbf{\bibinfo{volume}{39}}, \bibinfo{eid}{195014} (\bibinfo{year}{2022}),
  \eprint{2204.02190}.

\bibitem[{\citenamefont{{Belfiglio} et~al.}(2022)\citenamefont{{Belfiglio},
  {Giamb{\`o}}, and {Luongo}}}]{mio2022}
\bibinfo{author}{\bibfnamefont{A.}~\bibnamefont{{Belfiglio}}},
  \bibinfo{author}{\bibfnamefont{R.}~\bibnamefont{{Giamb{\`o}}}},
  \bibnamefont{and} \bibinfo{author}{\bibfnamefont{O.}~\bibnamefont{{Luongo}}},
  \bibinfo{journal}{arXiv e-prints} \bibinfo{eid}{arXiv:2206.14158}
  (\bibinfo{year}{2022}), \eprint{2206.14158}.

\bibitem[{\citenamefont{Hu and Wang}(2023)}]{Hu:2023jqc}
\bibinfo{author}{\bibfnamefont{J.-P.} \bibnamefont{Hu}} \bibnamefont{and}
  \bibinfo{author}{\bibfnamefont{F.-Y.} \bibnamefont{Wang}},
  \bibinfo{journal}{Universe} \textbf{\bibinfo{volume}{9}}, \bibinfo{pages}{94}
  (\bibinfo{year}{2023}), \eprint{2302.05709}.

\bibitem[{\citenamefont{{DESI Collaboration}}(2024)}]{2024arXiv240403002D}
\bibinfo{author}{\bibnamefont{{DESI Collaboration}}}, \bibinfo{journal}{arXiv
  e-prints} p. \bibinfo{pages}{arXiv:2404.03002} (\bibinfo{year}{2024}),
  \eprint{2404.03002}.

\bibitem[{\citenamefont{{Scolnic} et~al.}(2018)\citenamefont{{Scolnic},
  {Jones}, {Rest}, {Pan}, {Chornock}, {Foley}, {Huber}, {Kessler}, {Narayan},
  {Riess} et~al.}}]{2018ApJ...859..101S}
\bibinfo{author}{\bibfnamefont{D.~M.} \bibnamefont{{Scolnic}}},
  \bibinfo{author}{\bibfnamefont{D.~O.} \bibnamefont{{Jones}}},
  \bibinfo{author}{\bibfnamefont{A.}~\bibnamefont{{Rest}}},
  \bibinfo{author}{\bibfnamefont{Y.~C.} \bibnamefont{{Pan}}},
  \bibinfo{author}{\bibfnamefont{R.}~\bibnamefont{{Chornock}}},
  \bibinfo{author}{\bibfnamefont{R.~J.} \bibnamefont{{Foley}}},
  \bibinfo{author}{\bibfnamefont{M.~E.} \bibnamefont{{Huber}}},
  \bibinfo{author}{\bibfnamefont{R.}~\bibnamefont{{Kessler}}},
  \bibinfo{author}{\bibfnamefont{G.}~\bibnamefont{{Narayan}}},
  \bibinfo{author}{\bibfnamefont{A.~G.} \bibnamefont{{Riess}}},
  \bibnamefont{et~al.}, \bibinfo{journal}{\apj} \textbf{\bibinfo{volume}{859}},
  \bibinfo{eid}{101} (\bibinfo{year}{2018}), \eprint{1710.00845}.

\bibitem[{\citenamefont{{Cuceu} et~al.}(2019)\citenamefont{{Cuceu}, {Farr},
  {Lemos}, and {Font-Ribera}}}]{2019JCAP...10..044C}
\bibinfo{author}{\bibfnamefont{A.}~\bibnamefont{{Cuceu}}},
  \bibinfo{author}{\bibfnamefont{J.}~\bibnamefont{{Farr}}},
  \bibinfo{author}{\bibfnamefont{P.}~\bibnamefont{{Lemos}}}, \bibnamefont{and}
  \bibinfo{author}{\bibfnamefont{A.}~\bibnamefont{{Font-Ribera}}},
  \bibinfo{journal}{\jcap} \textbf{\bibinfo{volume}{2019}}, \bibinfo{eid}{044}
  (\bibinfo{year}{2019}), \eprint{1906.11628}.

\bibitem[{\citenamefont{{Rodney} et~al.}(2015)\citenamefont{{Rodney}, {Riess},
  {Scolnic}, {Jones}, {Hemmati}, {Molino}, {McCully}, {Mobasher}, {Strolger},
  {Graur} et~al.}}]{Rodney2015}
\bibinfo{author}{\bibfnamefont{S.~A.} \bibnamefont{{Rodney}}},
  \bibinfo{author}{\bibfnamefont{A.~G.} \bibnamefont{{Riess}}},
  \bibinfo{author}{\bibfnamefont{D.~M.} \bibnamefont{{Scolnic}}},
  \bibinfo{author}{\bibfnamefont{D.~O.} \bibnamefont{{Jones}}},
  \bibinfo{author}{\bibfnamefont{S.}~\bibnamefont{{Hemmati}}},
  \bibinfo{author}{\bibfnamefont{A.}~\bibnamefont{{Molino}}},
  \bibinfo{author}{\bibfnamefont{C.}~\bibnamefont{{McCully}}},
  \bibinfo{author}{\bibfnamefont{B.}~\bibnamefont{{Mobasher}}},
  \bibinfo{author}{\bibfnamefont{L.-G.} \bibnamefont{{Strolger}}},
  \bibinfo{author}{\bibfnamefont{O.}~\bibnamefont{{Graur}}},
  \bibnamefont{et~al.}, \bibinfo{journal}{\aj} \textbf{\bibinfo{volume}{150}},
  \bibinfo{eid}{156} (\bibinfo{year}{2015}).

\bibitem[{\citenamefont{{Capozziello} et~al.}(2019)\citenamefont{{Capozziello},
  {D'Agostino}, and {Luongo}}}]{2019IJMPD..2830016C}
\bibinfo{author}{\bibfnamefont{S.}~\bibnamefont{{Capozziello}}},
  \bibinfo{author}{\bibfnamefont{R.}~\bibnamefont{{D'Agostino}}},
  \bibnamefont{and} \bibinfo{author}{\bibfnamefont{O.}~\bibnamefont{{Luongo}}},
  \bibinfo{journal}{International Journal of Modern Physics D}
  \textbf{\bibinfo{volume}{28}}, \bibinfo{eid}{1930016} (\bibinfo{year}{2019}),
  \eprint{1904.01427}.

\bibitem[{\citenamefont{{Capozziello} et~al.}(2020)\citenamefont{{Capozziello},
  {D'Agostino}, and {Luongo}}}]{2020arXiv200309341C}
\bibinfo{author}{\bibfnamefont{S.}~\bibnamefont{{Capozziello}}},
  \bibinfo{author}{\bibfnamefont{R.}~\bibnamefont{{D'Agostino}}},
  \bibnamefont{and} \bibinfo{author}{\bibfnamefont{O.}~\bibnamefont{{Luongo}}},
  \bibinfo{journal}{\mnras} \textbf{\bibinfo{volume}{494}},
  \bibinfo{pages}{2576} (\bibinfo{year}{2020}), \eprint{2003.09341}.

\bibitem[{\citenamefont{{Muccino} et~al.}(2021)\citenamefont{{Muccino}, {Izzo},
  {Luongo}, {Boshkayev}, {Amati}, {Della Valle}, {Pisani}, and
  {Zaninoni}}}]{2021ApJ...908..181M}
\bibinfo{author}{\bibfnamefont{M.}~\bibnamefont{{Muccino}}},
  \bibinfo{author}{\bibfnamefont{L.}~\bibnamefont{{Izzo}}},
  \bibinfo{author}{\bibfnamefont{O.}~\bibnamefont{{Luongo}}},
  \bibinfo{author}{\bibfnamefont{K.}~\bibnamefont{{Boshkayev}}},
  \bibinfo{author}{\bibfnamefont{L.}~\bibnamefont{{Amati}}},
  \bibinfo{author}{\bibfnamefont{M.}~\bibnamefont{{Della Valle}}},
  \bibinfo{author}{\bibfnamefont{G.~B.} \bibnamefont{{Pisani}}},
  \bibnamefont{and}
  \bibinfo{author}{\bibfnamefont{E.}~\bibnamefont{{Zaninoni}}},
  \bibinfo{journal}{\apj} \textbf{\bibinfo{volume}{908}}, \bibinfo{eid}{181}
  (\bibinfo{year}{2021}), \eprint{2012.03392}.

\bibitem[{\citenamefont{{Cao} et~al.}(2022)\citenamefont{{Cao}, {Dainotti}, and
  {Ratra}}}]{2022MNRAS.512..439C}
\bibinfo{author}{\bibfnamefont{S.}~\bibnamefont{{Cao}}},
  \bibinfo{author}{\bibfnamefont{M.}~\bibnamefont{{Dainotti}}},
  \bibnamefont{and} \bibinfo{author}{\bibfnamefont{B.}~\bibnamefont{{Ratra}}},
  \bibinfo{journal}{\mnras} \textbf{\bibinfo{volume}{512}},
  \bibinfo{pages}{439} (\bibinfo{year}{2022}), \eprint{2201.05245}.

\bibitem[{\citenamefont{{Luongo} and
  {Muccino}}(2021{\natexlab{a}})}]{2021Galax...9...77L}
\bibinfo{author}{\bibfnamefont{O.}~\bibnamefont{{Luongo}}} \bibnamefont{and}
  \bibinfo{author}{\bibfnamefont{M.}~\bibnamefont{{Muccino}}},
  \bibinfo{journal}{Galaxies} \textbf{\bibinfo{volume}{9}}, \bibinfo{pages}{77}
  (\bibinfo{year}{2021}{\natexlab{a}}), \eprint{2110.14408}.

\bibitem[{\citenamefont{{Dainotti} et~al.}(2022)\citenamefont{{Dainotti},
  {Sarracino}, and {Capozziello}}}]{2022PASJ..tmp...83D}
\bibinfo{author}{\bibfnamefont{M.~G.} \bibnamefont{{Dainotti}}},
  \bibinfo{author}{\bibfnamefont{G.}~\bibnamefont{{Sarracino}}},
  \bibnamefont{and}
  \bibinfo{author}{\bibfnamefont{S.}~\bibnamefont{{Capozziello}}},
  \bibinfo{journal}{\pasj}  (\bibinfo{year}{2022}), \eprint{2206.07479}.

\bibitem[{\citenamefont{{Jia} et~al.}(2022)\citenamefont{{Jia}, {Hu}, {Yang},
  {Zhang}, and {Wang}}}]{2022arXiv220809272J}
\bibinfo{author}{\bibfnamefont{X.~D.} \bibnamefont{{Jia}}},
  \bibinfo{author}{\bibfnamefont{J.~P.} \bibnamefont{{Hu}}},
  \bibinfo{author}{\bibfnamefont{J.}~\bibnamefont{{Yang}}},
  \bibinfo{author}{\bibfnamefont{B.~B.} \bibnamefont{{Zhang}}},
  \bibnamefont{and} \bibinfo{author}{\bibfnamefont{F.~Y.}
  \bibnamefont{{Wang}}}, \bibinfo{journal}{arXiv e-prints}
  \bibinfo{eid}{arXiv:2208.09272} (\bibinfo{year}{2022}), \eprint{2208.09272}.

\bibitem[{\citenamefont{{Arjona} et~al.}(2019)\citenamefont{{Arjona},
  {Cardona}, and {Nesseris}}}]{2019PhRvD..99d3516A}
\bibinfo{author}{\bibfnamefont{R.}~\bibnamefont{{Arjona}}},
  \bibinfo{author}{\bibfnamefont{W.}~\bibnamefont{{Cardona}}},
  \bibnamefont{and}
  \bibinfo{author}{\bibfnamefont{S.}~\bibnamefont{{Nesseris}}},
  \bibinfo{journal}{\prd} \textbf{\bibinfo{volume}{99}}, \bibinfo{eid}{043516}
  (\bibinfo{year}{2019}), \eprint{1811.02469}.

\bibitem[{\citenamefont{{Amati} et~al.}(2019)\citenamefont{{Amati},
  {D'Agostino}, {Luongo}, {Muccino}, and {Tantalo}}}]{2019MNRAS.486L..46A}
\bibinfo{author}{\bibfnamefont{L.}~\bibnamefont{{Amati}}},
  \bibinfo{author}{\bibfnamefont{R.}~\bibnamefont{{D'Agostino}}},
  \bibinfo{author}{\bibfnamefont{O.}~\bibnamefont{{Luongo}}},
  \bibinfo{author}{\bibfnamefont{M.}~\bibnamefont{{Muccino}}},
  \bibnamefont{and}
  \bibinfo{author}{\bibfnamefont{M.}~\bibnamefont{{Tantalo}}},
  \bibinfo{journal}{\mnras} \textbf{\bibinfo{volume}{486}},
  \bibinfo{pages}{L46} (\bibinfo{year}{2019}), \eprint{1811.08934}.

\bibitem[{\citenamefont{{Luongo} and {Muccino}}(2021{\natexlab{b}})}]{LM2020}
\bibinfo{author}{\bibfnamefont{O.}~\bibnamefont{{Luongo}}} \bibnamefont{and}
  \bibinfo{author}{\bibfnamefont{M.}~\bibnamefont{{Muccino}}},
  \bibinfo{journal}{\mnras} \textbf{\bibinfo{volume}{503}},
  \bibinfo{pages}{4581} (\bibinfo{year}{2021}{\natexlab{b}}),
  \eprint{2011.13590}.

\bibitem[{\citenamefont{{Luongo} and {Muccino}}(2023)}]{2023MNRAS.518.2247L}
\bibinfo{author}{\bibfnamefont{O.}~\bibnamefont{{Luongo}}} \bibnamefont{and}
  \bibinfo{author}{\bibfnamefont{M.}~\bibnamefont{{Muccino}}},
  \bibinfo{journal}{\mnras} \textbf{\bibinfo{volume}{518}},
  \bibinfo{pages}{2247} (\bibinfo{year}{2023}), \eprint{2207.00440}.

\bibitem[{\citenamefont{{Muccino} et~al.}(2023)\citenamefont{{Muccino},
  {Luongo}, and {Jain}}}]{2023MNRAS.523.4938M}
\bibinfo{author}{\bibfnamefont{M.}~\bibnamefont{{Muccino}}},
  \bibinfo{author}{\bibfnamefont{O.}~\bibnamefont{{Luongo}}}, \bibnamefont{and}
  \bibinfo{author}{\bibfnamefont{D.}~\bibnamefont{{Jain}}},
  \bibinfo{journal}{\mnras} \textbf{\bibinfo{volume}{523}},
  \bibinfo{pages}{4938} (\bibinfo{year}{2023}), \eprint{2208.13700}.

\bibitem[{\citenamefont{{Alfano}
  et~al.}(2024{\natexlab{a}})\citenamefont{{Alfano}, {Luongo}, and
  {Muccino}}}]{2024arXiv241104878A}
\bibinfo{author}{\bibfnamefont{A.~C.} \bibnamefont{{Alfano}}},
  \bibinfo{author}{\bibfnamefont{O.}~\bibnamefont{{Luongo}}}, \bibnamefont{and}
  \bibinfo{author}{\bibfnamefont{M.}~\bibnamefont{{Muccino}}},
  \bibinfo{journal}{arXiv e-prints} \bibinfo{eid}{arXiv:2411.04878}
  (\bibinfo{year}{2024}{\natexlab{a}}), \eprint{2411.04878}.

\bibitem[{\citenamefont{{Alfano}
  et~al.}(2024{\natexlab{b}})\citenamefont{{Alfano}, {Luongo}, and
  {Muccino}}}]{2024arXiv240802536A}
\bibinfo{author}{\bibfnamefont{A.~C.} \bibnamefont{{Alfano}}},
  \bibinfo{author}{\bibfnamefont{O.}~\bibnamefont{{Luongo}}}, \bibnamefont{and}
  \bibinfo{author}{\bibfnamefont{M.}~\bibnamefont{{Muccino}}},
  \bibinfo{journal}{arXiv e-prints} \bibinfo{eid}{arXiv:2408.02536}
  (\bibinfo{year}{2024}{\natexlab{b}}), \eprint{2408.02536}.

\bibitem[{\citenamefont{{Carloni} et~al.}(2024)\citenamefont{{Carloni},
  {Luongo}, and {Muccino}}}]{2024arXiv240412068C}
\bibinfo{author}{\bibfnamefont{Y.}~\bibnamefont{{Carloni}}},
  \bibinfo{author}{\bibfnamefont{O.}~\bibnamefont{{Luongo}}}, \bibnamefont{and}
  \bibinfo{author}{\bibfnamefont{M.}~\bibnamefont{{Muccino}}},
  \bibinfo{journal}{arXiv e-prints} \bibinfo{eid}{arXiv:2404.12068}
  (\bibinfo{year}{2024}), \eprint{2404.12068}.

\bibitem[{\citenamefont{{Luongo} and {Muccino}}(2024)}]{2024A&A...690A..40L}
\bibinfo{author}{\bibfnamefont{O.}~\bibnamefont{{Luongo}}} \bibnamefont{and}
  \bibinfo{author}{\bibfnamefont{M.}~\bibnamefont{{Muccino}}},
  \bibinfo{journal}{\aap} \textbf{\bibinfo{volume}{690}}, \bibinfo{eid}{A40}
  (\bibinfo{year}{2024}), \eprint{2404.07070}.

\bibitem[{\citenamefont{Aviles and Cervantes-Cota}(2011)}]{Aviles:2011ak}
\bibinfo{author}{\bibfnamefont{A.}~\bibnamefont{Aviles}} \bibnamefont{and}
  \bibinfo{author}{\bibfnamefont{J.~L.} \bibnamefont{Cervantes-Cota}},
  \bibinfo{journal}{Phys. Rev. D} \textbf{\bibinfo{volume}{84}},
  \bibinfo{pages}{083515} (\bibinfo{year}{2011}), \bibinfo{note}{[Erratum:
  Phys.Rev.D 84, 089905 (2011)]}, \eprint{1108.2457}.

\bibitem[{\citenamefont{von Marttens et~al.}(2020)\citenamefont{von Marttens,
  Lombriser, Kunz, Marra, Casarini, and Alcaniz}}]{vonMarttens:2019ixw}
\bibinfo{author}{\bibfnamefont{R.}~\bibnamefont{von Marttens}},
  \bibinfo{author}{\bibfnamefont{L.}~\bibnamefont{Lombriser}},
  \bibinfo{author}{\bibfnamefont{M.}~\bibnamefont{Kunz}},
  \bibinfo{author}{\bibfnamefont{V.}~\bibnamefont{Marra}},
  \bibinfo{author}{\bibfnamefont{L.}~\bibnamefont{Casarini}}, \bibnamefont{and}
  \bibinfo{author}{\bibfnamefont{J.}~\bibnamefont{Alcaniz}},
  \bibinfo{journal}{Phys. Dark Univ.} \textbf{\bibinfo{volume}{28}},
  \bibinfo{pages}{100490} (\bibinfo{year}{2020}), \eprint{1911.02618}.

\bibitem[{\citenamefont{Luongo and Muccino}(2022)}]{Luongo:2022bju}
\bibinfo{author}{\bibfnamefont{O.}~\bibnamefont{Luongo}} \bibnamefont{and}
  \bibinfo{author}{\bibfnamefont{M.}~\bibnamefont{Muccino}},
  \bibinfo{journal}{Mon. Not. Roy. Astron. Soc.}
  \textbf{\bibinfo{volume}{518}}, \bibinfo{pages}{2247} (\bibinfo{year}{2022}),
  \eprint{2207.00440}.

\bibitem[{\citenamefont{Aviles et~al.}(2017)\citenamefont{Aviles, Klapp, and
  Luongo}}]{Aviles:2016wel}
\bibinfo{author}{\bibfnamefont{A.}~\bibnamefont{Aviles}},
  \bibinfo{author}{\bibfnamefont{J.}~\bibnamefont{Klapp}}, \bibnamefont{and}
  \bibinfo{author}{\bibfnamefont{O.}~\bibnamefont{Luongo}},
  \bibinfo{journal}{Phys. Dark Univ.} \textbf{\bibinfo{volume}{17}},
  \bibinfo{pages}{25} (\bibinfo{year}{2017}), \eprint{1606.09195}.

\bibitem[{\citenamefont{Izzo et~al.}(2012)\citenamefont{Izzo, Luongo, and
  Capozziello}}]{Izzo:2010ix}
\bibinfo{author}{\bibfnamefont{L.}~\bibnamefont{Izzo}},
  \bibinfo{author}{\bibfnamefont{O.}~\bibnamefont{Luongo}}, \bibnamefont{and}
  \bibinfo{author}{\bibfnamefont{S.}~\bibnamefont{Capozziello}},
  \bibinfo{journal}{Mem. Soc. Astron. Ital. Suppl.}
  \textbf{\bibinfo{volume}{19}}, \bibinfo{pages}{37} (\bibinfo{year}{2012}),
  \eprint{1011.1151}.

\bibitem[{\citenamefont{Aviles et~al.}(2012)\citenamefont{Aviles, Gruber,
  Luongo, and Quevedo}}]{Aviles:2012ay}
\bibinfo{author}{\bibfnamefont{A.}~\bibnamefont{Aviles}},
  \bibinfo{author}{\bibfnamefont{C.}~\bibnamefont{Gruber}},
  \bibinfo{author}{\bibfnamefont{O.}~\bibnamefont{Luongo}}, \bibnamefont{and}
  \bibinfo{author}{\bibfnamefont{H.}~\bibnamefont{Quevedo}},
  \bibinfo{journal}{Phys. Rev. D} \textbf{\bibinfo{volume}{86}},
  \bibinfo{pages}{123516} (\bibinfo{year}{2012}), \eprint{1204.2007}.

\bibitem[{\citenamefont{Dunsby and Luongo}(2016)}]{Dunsby:2015ers}
\bibinfo{author}{\bibfnamefont{P.~K.~S.} \bibnamefont{Dunsby}}
  \bibnamefont{and} \bibinfo{author}{\bibfnamefont{O.}~\bibnamefont{Luongo}},
  \bibinfo{journal}{Int. J. Geom. Meth. Mod. Phys.}
  \textbf{\bibinfo{volume}{13}}, \bibinfo{pages}{1630002}
  (\bibinfo{year}{2016}), \eprint{1511.06532}.

\bibitem[{\citenamefont{{Lobo} et~al.}(2020)\citenamefont{{Lobo}, {Mimoso}, and
  {Visser}}}]{2020JCAP...04..043L}
\bibinfo{author}{\bibfnamefont{F.~S.~N.} \bibnamefont{{Lobo}}},
  \bibinfo{author}{\bibfnamefont{J.~P.} \bibnamefont{{Mimoso}}},
  \bibnamefont{and} \bibinfo{author}{\bibfnamefont{M.}~\bibnamefont{{Visser}}},
  \bibinfo{journal}{\jcap} \textbf{\bibinfo{volume}{2020}}, \bibinfo{eid}{043}
  (\bibinfo{year}{2020}), \eprint{2001.11964}.

\bibitem[{\citenamefont{{Tucker} et~al.}(2005)\citenamefont{{Tucker}, {Burton},
  and {Noble}}}]{2005GReGr..37.1555T}
\bibinfo{author}{\bibfnamefont{R.~W.} \bibnamefont{{Tucker}}},
  \bibinfo{author}{\bibfnamefont{D.~A.} \bibnamefont{{Burton}}},
  \bibnamefont{and} \bibinfo{author}{\bibfnamefont{A.}~\bibnamefont{{Noble}}},
  \bibinfo{journal}{General Relativity and Gravitation}
  \textbf{\bibinfo{volume}{37}}, \bibinfo{pages}{1555} (\bibinfo{year}{2005}),
  \eprint{gr-qc/0411131}.

\bibitem[{\citenamefont{Luongo et~al.}(2016)\citenamefont{Luongo, Pisani, and
  Troisi}}]{Luongo:2015zgq}
\bibinfo{author}{\bibfnamefont{O.}~\bibnamefont{Luongo}},
  \bibinfo{author}{\bibfnamefont{G.~B.} \bibnamefont{Pisani}},
  \bibnamefont{and} \bibinfo{author}{\bibfnamefont{A.}~\bibnamefont{Troisi}},
  \bibinfo{journal}{Int. J. Mod. Phys. D} \textbf{\bibinfo{volume}{26}},
  \bibinfo{pages}{1750015} (\bibinfo{year}{2016}), \eprint{1512.07076}.

\bibitem[{\citenamefont{Capozziello et~al.}(2022)\citenamefont{Capozziello,
  D'Agostino, and Luongo}}]{Capozziello:2022jbw}
\bibinfo{author}{\bibfnamefont{S.}~\bibnamefont{Capozziello}},
  \bibinfo{author}{\bibfnamefont{R.}~\bibnamefont{D'Agostino}},
  \bibnamefont{and} \bibinfo{author}{\bibfnamefont{O.}~\bibnamefont{Luongo}},
  \bibinfo{journal}{Phys. Dark Univ.} \textbf{\bibinfo{volume}{36}},
  \bibinfo{pages}{101045} (\bibinfo{year}{2022}), \eprint{2202.03300}.

\bibitem[{\citenamefont{{Luongo} and
  {Muccino}}(2018{\natexlab{b}})}]{2018PhRvD..98j3520L}
\bibinfo{author}{\bibfnamefont{O.}~\bibnamefont{{Luongo}}} \bibnamefont{and}
  \bibinfo{author}{\bibfnamefont{M.}~\bibnamefont{{Muccino}}},
  \bibinfo{journal}{\prd} \textbf{\bibinfo{volume}{98}}, \bibinfo{eid}{103520}
  (\bibinfo{year}{2018}{\natexlab{b}}), \eprint{1807.00180}.

\bibitem[{\citenamefont{{Belfiglio}
  et~al.}(2024{\natexlab{a}})\citenamefont{{Belfiglio}, {Carloni}, and
  {Luongo}}}]{2024PDU....4401458B}
\bibinfo{author}{\bibfnamefont{A.}~\bibnamefont{{Belfiglio}}},
  \bibinfo{author}{\bibfnamefont{Y.}~\bibnamefont{{Carloni}}},
  \bibnamefont{and} \bibinfo{author}{\bibfnamefont{O.}~\bibnamefont{{Luongo}}},
  \bibinfo{journal}{Physics of the Dark Universe}
  \textbf{\bibinfo{volume}{44}}, \bibinfo{eid}{101458}
  (\bibinfo{year}{2024}{\natexlab{a}}), \eprint{2307.04739}.

\bibitem[{\citenamefont{{Belfiglio}
  et~al.}(2024{\natexlab{b}})\citenamefont{{Belfiglio}, {Luongo}, and
  {Mengoni}}}]{2024arXiv241111130B}
\bibinfo{author}{\bibfnamefont{A.}~\bibnamefont{{Belfiglio}}},
  \bibinfo{author}{\bibfnamefont{O.}~\bibnamefont{{Luongo}}}, \bibnamefont{and}
  \bibinfo{author}{\bibfnamefont{T.}~\bibnamefont{{Mengoni}}},
  \bibinfo{journal}{arXiv e-prints} \bibinfo{eid}{arXiv:2411.11130}
  (\bibinfo{year}{2024}{\natexlab{b}}), \eprint{2411.11130}.

\bibitem[{\citenamefont{{Alfano} et~al.}(2023)\citenamefont{{Alfano}, {Cafaro},
  {Capozziello}, and {Luongo}}}]{2023PDU....4201298A}
\bibinfo{author}{\bibfnamefont{A.~C.} \bibnamefont{{Alfano}}},
  \bibinfo{author}{\bibfnamefont{C.}~\bibnamefont{{Cafaro}}},
  \bibinfo{author}{\bibfnamefont{S.}~\bibnamefont{{Capozziello}}},
  \bibnamefont{and} \bibinfo{author}{\bibfnamefont{O.}~\bibnamefont{{Luongo}}},
  \bibinfo{journal}{Physics of the Dark Universe}
  \textbf{\bibinfo{volume}{42}}, \bibinfo{eid}{101298} (\bibinfo{year}{2023}),
  \eprint{2306.08396}.

\bibitem[{\citenamefont{{Melchiorri} et~al.}(2007)\citenamefont{{Melchiorri},
  {Pagano}, and {Pandolfi}}}]{2007PhRvD..76d1301M}
\bibinfo{author}{\bibfnamefont{A.}~\bibnamefont{{Melchiorri}}},
  \bibinfo{author}{\bibfnamefont{L.}~\bibnamefont{{Pagano}}}, \bibnamefont{and}
  \bibinfo{author}{\bibfnamefont{S.}~\bibnamefont{{Pandolfi}}},
  \bibinfo{journal}{\prd} \textbf{\bibinfo{volume}{76}}, \bibinfo{eid}{041301}
  (\bibinfo{year}{2007}), \eprint{0706.1314}.

\bibitem[{\citenamefont{{Alfano}
  et~al.}(2024{\natexlab{c}})\citenamefont{{Alfano}, {Capozziello}, {Luongo},
  and {Muccino}}}]{2024JHEAp..42..178A}
\bibinfo{author}{\bibfnamefont{A.~C.} \bibnamefont{{Alfano}}},
  \bibinfo{author}{\bibfnamefont{S.}~\bibnamefont{{Capozziello}}},
  \bibinfo{author}{\bibfnamefont{O.}~\bibnamefont{{Luongo}}}, \bibnamefont{and}
  \bibinfo{author}{\bibfnamefont{M.}~\bibnamefont{{Muccino}}},
  \bibinfo{journal}{Journal of High Energy Astrophysics}
  \textbf{\bibinfo{volume}{42}}, \bibinfo{pages}{178}
  (\bibinfo{year}{2024}{\natexlab{c}}), \eprint{2402.18967}.

\bibitem[{\citenamefont{{Capozziello} et~al.}(2022)\citenamefont{{Capozziello},
  {Dunsby}, and {Luongo}}}]{2022MNRAS.509.5399C}
\bibinfo{author}{\bibfnamefont{S.}~\bibnamefont{{Capozziello}}},
  \bibinfo{author}{\bibfnamefont{P.~K.~S.} \bibnamefont{{Dunsby}}},
  \bibnamefont{and} \bibinfo{author}{\bibfnamefont{O.}~\bibnamefont{{Luongo}}},
  \bibinfo{journal}{\mnras} \textbf{\bibinfo{volume}{509}},
  \bibinfo{pages}{5399} (\bibinfo{year}{2022}), \eprint{2106.15579}.

\bibitem[{\citenamefont{{Dunsby} and {Luongo}}(2016)}]{2016IJGMM..1330002D}
\bibinfo{author}{\bibfnamefont{P.~K.~S.} \bibnamefont{{Dunsby}}}
  \bibnamefont{and} \bibinfo{author}{\bibfnamefont{O.}~\bibnamefont{{Luongo}}},
  \bibinfo{journal}{International Journal of Geometric Methods in Modern
  Physics} \textbf{\bibinfo{volume}{13}}, \bibinfo{eid}{1630002-606}
  (\bibinfo{year}{2016}), \eprint{1511.06532}.

\bibitem[{\citenamefont{{Visser}}(2005)}]{visser}
\bibinfo{author}{\bibfnamefont{M.}~\bibnamefont{{Visser}}},
  \bibinfo{journal}{Gen. Rev. Grav.} \textbf{\bibinfo{volume}{37}},
  \bibinfo{eid}{1541} (\bibinfo{year}{2005}), \eprint{0411131}.

\bibitem[{\citenamefont{{Visser} et~al.}(2010)\citenamefont{{Visser},
  {Catto{\"e}n}, and {}}}]{2010dmap.conf..287V}
\bibinfo{author}{\bibfnamefont{M.}~\bibnamefont{{Visser}}},
  \bibinfo{author}{\bibnamefont{{Catto{\"e}n}}}, \bibnamefont{and}
  \bibinfo{author}{\bibfnamefont{C.}~\bibnamefont{{}}}, in
  \emph{\bibinfo{booktitle}{Dark Matter in Astrophysics and Particle Physics,
  Dark 2009}}, edited by \bibinfo{editor}{\bibfnamefont{H.~V.}
  \bibnamefont{{Klapdor-Kleingrothaus}}} \bibnamefont{and}
  \bibinfo{editor}{\bibfnamefont{I.~V.} \bibnamefont{{Krivosheina}}}
  (\bibinfo{year}{2010}), pp. \bibinfo{pages}{287--300}, \eprint{0906.5407}.

\bibitem[{\citenamefont{{Catto{\"e}n} and
  {Visser}}(2007)}]{2007CQGra..24.5985C}
\bibinfo{author}{\bibfnamefont{C.}~\bibnamefont{{Catto{\"e}n}}}
  \bibnamefont{and} \bibinfo{author}{\bibfnamefont{M.}~\bibnamefont{{Visser}}},
  \bibinfo{journal}{Classical and Quantum Gravity}
  \textbf{\bibinfo{volume}{24}}, \bibinfo{pages}{5985} (\bibinfo{year}{2007}),
  \eprint{0710.1887}.

\bibitem[{\citenamefont{Muccino et~al.}(2021)\citenamefont{Muccino, Izzo,
  Luongo, Boshkayev, Amati, Della~Valle, Pisani, and
  Zaninoni}}]{Muccino:2020gqt}
\bibinfo{author}{\bibfnamefont{M.}~\bibnamefont{Muccino}},
  \bibinfo{author}{\bibfnamefont{L.}~\bibnamefont{Izzo}},
  \bibinfo{author}{\bibfnamefont{O.}~\bibnamefont{Luongo}},
  \bibinfo{author}{\bibfnamefont{K.}~\bibnamefont{Boshkayev}},
  \bibinfo{author}{\bibfnamefont{L.}~\bibnamefont{Amati}},
  \bibinfo{author}{\bibfnamefont{M.}~\bibnamefont{Della~Valle}},
  \bibinfo{author}{\bibfnamefont{G.~B.} \bibnamefont{Pisani}},
  \bibnamefont{and} \bibinfo{author}{\bibfnamefont{E.}~\bibnamefont{Zaninoni}},
  \bibinfo{journal}{Astrophys. J.} \textbf{\bibinfo{volume}{908}},
  \bibinfo{pages}{181} (\bibinfo{year}{2021}), \eprint{2012.03392}.

\bibitem[{\citenamefont{{Martins} and {Prat
  Colomer}}(2018)}]{2018A&A...616A..32M}
\bibinfo{author}{\bibfnamefont{C.~J.~A.~P.} \bibnamefont{{Martins}}}
  \bibnamefont{and} \bibinfo{author}{\bibfnamefont{M.}~\bibnamefont{{Prat
  Colomer}}}, \bibinfo{journal}{\aap} \textbf{\bibinfo{volume}{616}},
  \bibinfo{eid}{A32} (\bibinfo{year}{2018}), \eprint{1806.07653}.

\bibitem[{\citenamefont{Li et~al.}(2013)\citenamefont{Li, Li, Wang, and
  Wang}}]{Li:2012dt}
\bibinfo{author}{\bibfnamefont{M.}~\bibnamefont{Li}},
  \bibinfo{author}{\bibfnamefont{X.-D.} \bibnamefont{Li}},
  \bibinfo{author}{\bibfnamefont{S.}~\bibnamefont{Wang}}, \bibnamefont{and}
  \bibinfo{author}{\bibfnamefont{Y.}~\bibnamefont{Wang}},
  \bibinfo{journal}{Front. Phys. (Beijing)} \textbf{\bibinfo{volume}{8}},
  \bibinfo{pages}{828} (\bibinfo{year}{2013}), \eprint{1209.0922}.

\bibitem[{\citenamefont{{Kumar} et~al.}(2023)\citenamefont{{Kumar}, {Jain},
  {Mahajan}, {Mukherjee}, and {Rana}}}]{2023IJMPD..3250039K}
\bibinfo{author}{\bibfnamefont{D.}~\bibnamefont{{Kumar}}},
  \bibinfo{author}{\bibfnamefont{D.}~\bibnamefont{{Jain}}},
  \bibinfo{author}{\bibfnamefont{S.}~\bibnamefont{{Mahajan}}},
  \bibinfo{author}{\bibfnamefont{A.}~\bibnamefont{{Mukherjee}}},
  \bibnamefont{and} \bibinfo{author}{\bibfnamefont{A.}~\bibnamefont{{Rana}}},
  \bibinfo{journal}{International Journal of Modern Physics D}
  \textbf{\bibinfo{volume}{32}}, \bibinfo{eid}{2350039} (\bibinfo{year}{2023}),
  \eprint{2205.13247}.

\bibitem[{\citenamefont{{Rubano} and {Scudellaro}}(2002)}]{2002astro.ph..3225R}
\bibinfo{author}{\bibfnamefont{C.}~\bibnamefont{{Rubano}}} \bibnamefont{and}
  \bibinfo{author}{\bibfnamefont{P.}~\bibnamefont{{Scudellaro}}},
  \bibinfo{journal}{arXiv e-prints} \bibinfo{eid}{astro-ph/0203225}
  (\bibinfo{year}{2002}), \eprint{astro-ph/0203225}.

\bibitem[{\citenamefont{Luongo}(2025)}]{Orlando}
\bibinfo{author}{\bibfnamefont{O.}~\bibnamefont{Luongo}}, \bibinfo{journal}{In
  preparation}  (\bibinfo{year}{2025}).

\bibitem[{\citenamefont{Luongo and Muccino}(2021)}]{Luongo:2021pjs}
\bibinfo{author}{\bibfnamefont{O.}~\bibnamefont{Luongo}} \bibnamefont{and}
  \bibinfo{author}{\bibfnamefont{M.}~\bibnamefont{Muccino}},
  \bibinfo{journal}{Galaxies} \textbf{\bibinfo{volume}{9}}, \bibinfo{pages}{77}
  (\bibinfo{year}{2021}), \eprint{2110.14408}.

\bibitem[{\citenamefont{{Tanvir} et~al.}(2009)\citenamefont{{Tanvir}, {Fox},
  {Levan}, {Berger}, {Wiersema}, {Fynbo}, {Cucchiara}, {Kr{\"u}hler},
  {Gehrels}, {Bloom} et~al.}}]{Tanvir2009}
\bibinfo{author}{\bibfnamefont{N.~R.} \bibnamefont{{Tanvir}}},
  \bibinfo{author}{\bibfnamefont{D.~B.} \bibnamefont{{Fox}}},
  \bibinfo{author}{\bibfnamefont{A.~J.} \bibnamefont{{Levan}}},
  \bibinfo{author}{\bibfnamefont{E.}~\bibnamefont{{Berger}}},
  \bibinfo{author}{\bibfnamefont{K.}~\bibnamefont{{Wiersema}}},
  \bibinfo{author}{\bibfnamefont{J.~P.~U.} \bibnamefont{{Fynbo}}},
  \bibinfo{author}{\bibfnamefont{A.}~\bibnamefont{{Cucchiara}}},
  \bibinfo{author}{\bibfnamefont{T.}~\bibnamefont{{Kr{\"u}hler}}},
  \bibinfo{author}{\bibfnamefont{N.}~\bibnamefont{{Gehrels}}},
  \bibinfo{author}{\bibfnamefont{J.~S.} \bibnamefont{{Bloom}}},
  \bibnamefont{et~al.}, \bibinfo{journal}{\nat} \textbf{\bibinfo{volume}{461}},
  \bibinfo{pages}{1254} (\bibinfo{year}{2009}), \eprint{0906.1577}.

\bibitem[{\citenamefont{{Cucchiara} et~al.}(2011)\citenamefont{{Cucchiara},
  {Levan}, {Fox}, {Tanvir}, {Ukwatta}, {Berger}, {Kr{\"u}hler}, {K{\"u}pc{\"u}
  Yolda{\c s}}, {Wu}, {Toma} et~al.}}]{Cucchiara2011}
\bibinfo{author}{\bibfnamefont{A.}~\bibnamefont{{Cucchiara}}},
  \bibinfo{author}{\bibfnamefont{A.~J.} \bibnamefont{{Levan}}},
  \bibinfo{author}{\bibfnamefont{D.~B.} \bibnamefont{{Fox}}},
  \bibinfo{author}{\bibfnamefont{N.~R.} \bibnamefont{{Tanvir}}},
  \bibinfo{author}{\bibfnamefont{T.~N.} \bibnamefont{{Ukwatta}}},
  \bibinfo{author}{\bibfnamefont{E.}~\bibnamefont{{Berger}}},
  \bibinfo{author}{\bibfnamefont{T.}~\bibnamefont{{Kr{\"u}hler}}},
  \bibinfo{author}{\bibfnamefont{A.}~\bibnamefont{{K{\"u}pc{\"u} Yolda{\c
  s}}}}, \bibinfo{author}{\bibfnamefont{X.~F.} \bibnamefont{{Wu}}},
  \bibinfo{author}{\bibfnamefont{K.}~\bibnamefont{{Toma}}},
  \bibnamefont{et~al.}, \bibinfo{journal}{\apj} \textbf{\bibinfo{volume}{736}},
  \bibinfo{eid}{7} (\bibinfo{year}{2011}), \eprint{1105.4915}.

\bibitem[{\citenamefont{{Amati} et~al.}(2021)\citenamefont{{Amati}, {O'Brien},
  {G{\"o}tz}, {Bozzo}, {Santangelo}, {Tanvir}, {Frontera}, {Mereghetti},
  {Osborne}, {Blain} et~al.}}]{2021ExA....52..183A}
\bibinfo{author}{\bibfnamefont{L.}~\bibnamefont{{Amati}}},
  \bibinfo{author}{\bibfnamefont{P.~T.} \bibnamefont{{O'Brien}}},
  \bibinfo{author}{\bibfnamefont{D.}~\bibnamefont{{G{\"o}tz}}},
  \bibinfo{author}{\bibfnamefont{E.}~\bibnamefont{{Bozzo}}},
  \bibinfo{author}{\bibfnamefont{A.}~\bibnamefont{{Santangelo}}},
  \bibinfo{author}{\bibfnamefont{N.}~\bibnamefont{{Tanvir}}},
  \bibinfo{author}{\bibfnamefont{F.}~\bibnamefont{{Frontera}}},
  \bibinfo{author}{\bibfnamefont{S.}~\bibnamefont{{Mereghetti}}},
  \bibinfo{author}{\bibfnamefont{J.~P.} \bibnamefont{{Osborne}}},
  \bibinfo{author}{\bibfnamefont{A.}~\bibnamefont{{Blain}}},
  \bibnamefont{et~al.}, \bibinfo{journal}{Experimental Astronomy}
  \textbf{\bibinfo{volume}{52}}, \bibinfo{pages}{183} (\bibinfo{year}{2021}),
  \eprint{2104.09531}.

\bibitem[{\citenamefont{{Amati} et~al.}(2002)\citenamefont{{Amati}, {Frontera},
  {Tavani}, {in't Zand}, {Antonelli}, {Costa}, {Feroci}, {Guidorzi}, {Heise},
  {Masetti} et~al.}}]{2002A&A...390...81A}
\bibinfo{author}{\bibfnamefont{L.}~\bibnamefont{{Amati}}},
  \bibinfo{author}{\bibfnamefont{F.}~\bibnamefont{{Frontera}}},
  \bibinfo{author}{\bibfnamefont{M.}~\bibnamefont{{Tavani}}},
  \bibinfo{author}{\bibfnamefont{J.~J.~M.} \bibnamefont{{in't Zand}}},
  \bibinfo{author}{\bibfnamefont{A.}~\bibnamefont{{Antonelli}}},
  \bibinfo{author}{\bibfnamefont{E.}~\bibnamefont{{Costa}}},
  \bibinfo{author}{\bibfnamefont{M.}~\bibnamefont{{Feroci}}},
  \bibinfo{author}{\bibfnamefont{C.}~\bibnamefont{{Guidorzi}}},
  \bibinfo{author}{\bibfnamefont{J.}~\bibnamefont{{Heise}}},
  \bibinfo{author}{\bibfnamefont{N.}~\bibnamefont{{Masetti}}},
  \bibnamefont{et~al.}, \bibinfo{journal}{\aap} \textbf{\bibinfo{volume}{390}},
  \bibinfo{pages}{81} (\bibinfo{year}{2002}), \eprint{astro-ph/0205230}.

\bibitem[{\citenamefont{{Amati} and {Della Valle}}(2013)}]{AmatiDellaValle2013}
\bibinfo{author}{\bibfnamefont{L.}~\bibnamefont{{Amati}}} \bibnamefont{and}
  \bibinfo{author}{\bibfnamefont{M.}~\bibnamefont{{Della Valle}}},
  \bibinfo{journal}{International Journal of Modern Physics D}
  \textbf{\bibinfo{volume}{22}}, \bibinfo{eid}{1330028} (\bibinfo{year}{2013}),
  \eprint{1310.3141}.

\bibitem[{\citenamefont{{Khadka} et~al.}(2021)\citenamefont{{Khadka}, {Luongo},
  {Muccino}, and {Ratra}}}]{2021JCAP...09..042K}
\bibinfo{author}{\bibfnamefont{N.}~\bibnamefont{{Khadka}}},
  \bibinfo{author}{\bibfnamefont{O.}~\bibnamefont{{Luongo}}},
  \bibinfo{author}{\bibfnamefont{M.}~\bibnamefont{{Muccino}}},
  \bibnamefont{and} \bibinfo{author}{\bibfnamefont{B.}~\bibnamefont{{Ratra}}},
  \bibinfo{journal}{\jcap} \textbf{\bibinfo{volume}{2021}}, \bibinfo{eid}{042}
  (\bibinfo{year}{2021}), \eprint{2105.12692}.

\bibitem[{\citenamefont{{Risaliti} and {Lusso}}(2019)}]{2019NatAs...3..272R}
\bibinfo{author}{\bibfnamefont{G.}~\bibnamefont{{Risaliti}}} \bibnamefont{and}
  \bibinfo{author}{\bibfnamefont{E.}~\bibnamefont{{Lusso}}},
  \bibinfo{journal}{Nature Astronomy} \textbf{\bibinfo{volume}{3}},
  \bibinfo{pages}{272} (\bibinfo{year}{2019}), \eprint{1811.02590}.

\bibitem[{\citenamefont{{Li}}(2007)}]{2007MNRAS.379L..55L}
\bibinfo{author}{\bibfnamefont{L.-X.} \bibnamefont{{Li}}},
  \bibinfo{journal}{\mnras} \textbf{\bibinfo{volume}{379}},
  \bibinfo{pages}{L55} (\bibinfo{year}{2007}), \eprint{0704.3128}.

\bibitem[{\citenamefont{{Kodama} et~al.}(2008)\citenamefont{{Kodama},
  {Yonetoku}, {Murakami}, {Tanabe}, {Tsutsui}, and {Nakamura}}}]{Kodama2008}
\bibinfo{author}{\bibfnamefont{Y.}~\bibnamefont{{Kodama}}},
  \bibinfo{author}{\bibfnamefont{D.}~\bibnamefont{{Yonetoku}}},
  \bibinfo{author}{\bibfnamefont{T.}~\bibnamefont{{Murakami}}},
  \bibinfo{author}{\bibfnamefont{S.}~\bibnamefont{{Tanabe}}},
  \bibinfo{author}{\bibfnamefont{R.}~\bibnamefont{{Tsutsui}}},
  \bibnamefont{and}
  \bibinfo{author}{\bibfnamefont{T.}~\bibnamefont{{Nakamura}}},
  \bibinfo{journal}{\mnras} \textbf{\bibinfo{volume}{391}}, \bibinfo{pages}{L1}
  (\bibinfo{year}{2008}), \eprint{0802.3428}.

\bibitem[{\citenamefont{{Montiel} et~al.}(2021)\citenamefont{{Montiel},
  {Cabrera}, and {Hidalgo}}}]{2021MNRAS.501.3515M}
\bibinfo{author}{\bibfnamefont{A.}~\bibnamefont{{Montiel}}},
  \bibinfo{author}{\bibfnamefont{J.~I.} \bibnamefont{{Cabrera}}},
  \bibnamefont{and} \bibinfo{author}{\bibfnamefont{J.~C.}
  \bibnamefont{{Hidalgo}}}, \bibinfo{journal}{\mnras}
  \textbf{\bibinfo{volume}{501}}, \bibinfo{pages}{3515} (\bibinfo{year}{2021}),
  \eprint{2003.03387}.

\bibitem[{\citenamefont{{Jimenez} and {Loeb}}(2002)}]{2002ApJ...573...37J}
\bibinfo{author}{\bibfnamefont{R.}~\bibnamefont{{Jimenez}}} \bibnamefont{and}
  \bibinfo{author}{\bibfnamefont{A.}~\bibnamefont{{Loeb}}},
  \bibinfo{journal}{\apj} \textbf{\bibinfo{volume}{573}}, \bibinfo{pages}{37}
  (\bibinfo{year}{2002}), \eprint{astro-ph/0106145}.

\bibitem[{\citenamefont{{Moresco} et~al.}(2020)\citenamefont{{Moresco},
  {Jimenez}, {Verde}, {Cimatti}, and {Pozzetti}}}]{2020ApJ...898...82M}
\bibinfo{author}{\bibfnamefont{M.}~\bibnamefont{{Moresco}}},
  \bibinfo{author}{\bibfnamefont{R.}~\bibnamefont{{Jimenez}}},
  \bibinfo{author}{\bibfnamefont{L.}~\bibnamefont{{Verde}}},
  \bibinfo{author}{\bibfnamefont{A.}~\bibnamefont{{Cimatti}}},
  \bibnamefont{and}
  \bibinfo{author}{\bibfnamefont{L.}~\bibnamefont{{Pozzetti}}},
  \bibinfo{journal}{\apj} \textbf{\bibinfo{volume}{898}}, \bibinfo{eid}{82}
  (\bibinfo{year}{2020}), \eprint{2003.07362}.

\bibitem[{\citenamefont{{Moresco} et~al.}(2022)\citenamefont{{Moresco},
  {Amati}, {Amendola}, {Birrer}, {Blakeslee}, {Cantiello}, {Cimatti},
  {Darling}, {Della Valle}, {Fishbach} et~al.}}]{2022LRR....25....6M}
\bibinfo{author}{\bibfnamefont{M.}~\bibnamefont{{Moresco}}},
  \bibinfo{author}{\bibfnamefont{L.}~\bibnamefont{{Amati}}},
  \bibinfo{author}{\bibfnamefont{L.}~\bibnamefont{{Amendola}}},
  \bibinfo{author}{\bibfnamefont{S.}~\bibnamefont{{Birrer}}},
  \bibinfo{author}{\bibfnamefont{J.~P.} \bibnamefont{{Blakeslee}}},
  \bibinfo{author}{\bibfnamefont{M.}~\bibnamefont{{Cantiello}}},
  \bibinfo{author}{\bibfnamefont{A.}~\bibnamefont{{Cimatti}}},
  \bibinfo{author}{\bibfnamefont{J.}~\bibnamefont{{Darling}}},
  \bibinfo{author}{\bibfnamefont{M.}~\bibnamefont{{Della Valle}}},
  \bibinfo{author}{\bibfnamefont{M.}~\bibnamefont{{Fishbach}}},
  \bibnamefont{et~al.}, \bibinfo{journal}{Living Reviews in Relativity}
  \textbf{\bibinfo{volume}{25}}, \bibinfo{eid}{6} (\bibinfo{year}{2022}),
  \eprint{2201.07241}.

\bibitem[{\citenamefont{{Riess} et~al.}(2022)\citenamefont{{Riess}, {Yuan},
  {Macri}, {Scolnic}, , {Brout} et~al.}}]{2022ApJ...934L...7R}
\bibinfo{author}{\bibfnamefont{A.~G.} \bibnamefont{{Riess}}},
  \bibinfo{author}{\bibfnamefont{W.}~\bibnamefont{{Yuan}}},
  \bibinfo{author}{\bibfnamefont{L.~M.} \bibnamefont{{Macri}}},
  \bibinfo{author}{\bibfnamefont{D.}~\bibnamefont{{Scolnic}}}, ,
  \bibinfo{author}{\bibfnamefont{D.}~\bibnamefont{{Brout}}},
  \bibnamefont{et~al.}, \bibinfo{journal}{\apjl}
  \textbf{\bibinfo{volume}{934}}, \bibinfo{eid}{L7} (\bibinfo{year}{2022}),
  \eprint{2112.04510}.

\bibitem[{\citenamefont{{Khetan} et~al.}(2021)\citenamefont{{Khetan}, {Izzo},
  {Branchesi}, {Wojtak}, {Cantiello}, {Murugeshan}, {Agnello}, {Cappellaro},
  {Della Valle}, {Gall} et~al.}}]{2021A&A...647A..72K}
\bibinfo{author}{\bibfnamefont{N.}~\bibnamefont{{Khetan}}},
  \bibinfo{author}{\bibfnamefont{L.}~\bibnamefont{{Izzo}}},
  \bibinfo{author}{\bibfnamefont{M.}~\bibnamefont{{Branchesi}}},
  \bibinfo{author}{\bibfnamefont{R.}~\bibnamefont{{Wojtak}}},
  \bibinfo{author}{\bibfnamefont{M.}~\bibnamefont{{Cantiello}}},
  \bibinfo{author}{\bibfnamefont{C.}~\bibnamefont{{Murugeshan}}},
  \bibinfo{author}{\bibfnamefont{A.}~\bibnamefont{{Agnello}}},
  \bibinfo{author}{\bibfnamefont{E.}~\bibnamefont{{Cappellaro}}},
  \bibinfo{author}{\bibfnamefont{M.}~\bibnamefont{{Della Valle}}},
  \bibinfo{author}{\bibfnamefont{C.}~\bibnamefont{{Gall}}},
  \bibnamefont{et~al.}, \bibinfo{journal}{\aap} \textbf{\bibinfo{volume}{647}},
  \bibinfo{eid}{A72} (\bibinfo{year}{2021}), \eprint{2008.07754}.

\end{thebibliography}
\end{document}